\documentclass[11pt,a4paper]{article}
\pdfoutput=1
\usepackage{jheppub}
\usepackage{amsthm}

\usepackage{subfigure} 
\usepackage{mathrsfs}
\usepackage{float}
\newcommand{\im}[1]{\text{Im}\left(#1\right)}

\newcommand{\VL}{sUVL\ }
\newcommand{\CKM}{V}
\newcommand{\Vfig}[1]{\CKM_{#1}}
\newcommand{\V}[1]{\CKM_{#1}^{\phantom{\ast}}}
\newcommand{\Vc}[1]{\CKM_{#1}^{\ast}}

\newcommand{\asld}{A^d_{SL}}
\newcommand{\asls}{A^s_{SL}}
\newcommand{\aslb}{A^b_{SL}}
\newcommand{\asldiff}{\asls-\asld}
\newcommand{\AJPKs}{A_{J/\Psi K_S}}
\newcommand{\AJPP}{A_{J/\Psi \Phi}}
\newcommand{\BBdmix}{$B^0_d$--$\bar B^0_d$}
\newcommand{\BBsmix}{$B^0_s$--$\bar B^0_s$}
\newcommand{\DDmix}{$D^0$--$\bar D^0$}
\newcommand{\KKmix}{$K^0$--$\bar K^0$}
\newcommand{\mT}{m_T}
\newcommand{\BTNu}{\text{Br}(B^+\to\tau^+\nu_\tau)}
\newcommand{\BsG}{\text{Br}(B\to X_s\gamma)}

\newcommand{\BXsmm}{\text{Br}(B\to X_s\mu^+\mu^-)}
\newcommand{\xD}{x_D}
\newcommand{\Dmm}{\text{Br}(D^0\to\mu^+\mu^-)}
\newcommand{\Bdmm}{\text{Br}(B_d\to\mu^+\mu^-)}
\newcommand{\Bsmm}{\text{Br}(B_s\to\mu^+\mu^-)}
\newcommand{\tcZ}{\text{Br}(t\to cZ)}
\newcommand{\tuZ}{\text{Br}(t\to uZ)}
\newcommand{\KPinunu}{\text{Br}(K^+\to\pi^+\nu\bar\nu)}
\newcommand{\KLPinunu}{\text{Br}(K_L\to\pi^0\nu\bar\nu)}

\newcommand{\DMBd}{\Delta M_{B_d}}

\newcommand{\mixMBd}{M_{12}^{B_d}}
\newcommand{\mixGBd}{\Gamma_{12}^{B_d}}

\newcommand{\mixMBq}{M_{12}^{B_q}}

\graphicspath{{Figs-v2/}}

\title{The Hunt for New Physics in the Flavour Sector with up vector-like quarks}

\author[a]{F. J. Botella,}
\author[b]{G. C. Branco}
\author[a]{and M. Nebot} 
\affiliation[a]{Departament de F\' \i sica Te\`orica and IFIC,\\ Universitat de Val\`encia - CSIC, E-46100, Burjassot, Spain}
\affiliation[b]{Departamento de F\'\i sica and Centro de F\' \i sica Te\' orica de Part\' \i culas,\\ Instituto Superior T\' ecnico, Universidade T\'ecnica de Lisboa, Av. Rovisco Pais, P-1049-001 Lisboa}

\emailAdd{fbotella@uv.es}
\emailAdd{gbranco@ist.utl.pt}
\emailAdd{negomi@uv.es}

\abstract{%
We analyse the possible presence of New Physics (NP) in the Flavour Sector and evaluate its potential for solving the tension between the experimental values of $\AJPKs$ and $\BTNu$ with respect 
to the Standard Model (SM) expectations. Updated model independent analyses, where
NP contributions are allowed in \BBdmix\ and \BBsmix\ transitions, suggest the need of New Physics in the $bd$ sector. A detailed analysis of recent Flavour data is then presented in the framework of a simple extension of the SM, where a $Q=2/3$
vector-like isosinglet quark is added to the spectrum of the SM. Special emphasis is given to the implications of this model for correlations among
various measurable quantities. We include constraints from all the relevant
quark flavour sectors and give precise predictions for selected rare processes. We find important deviations from the SM in observables in the $bd$
sector like the semileptonic asymmetry $\asld$, $B_{d}^{0}\to \mu ^{+}\mu ^{-}$ and $\asldiff$. 
Other potential places where NP can show up include $\AJPP$, $\gamma$, $K_{L}^{0}\to \pi^{0}\nu \bar{\nu}$, $t\to Zq$ and $D^{0}\to \mu ^{+}\mu ^{-}$ among others. The experimental data favours in this model the existence of an up
vector-like quark with a mass below $600(1000)$ GeV at $1(2)$ $\sigma $.%
}
\arxivnumber{1207.4440}

\begin{document}
\maketitle

%\newpage
% % % % % % % % % % % % % % % % %
\section{Introduction\label{sec:introduction}}
% % % % % % % % % % % % % % % % %

The flavour puzzle remains one of the fundamental questions in particle physics \cite{Buras:2011fz}. In the Standard Model (SM), the flavour structure of Yukawa couplings is not constrained by gauge invariance, which in turn implies that quark masses and their mixing are free parameters of the theory. Apart from this important shortcoming, the SM has been very successful in describing flavour mixing and CP violation through a $3\times 3$ unitary Cabibbo-Kobayashi-Maskawa (CKM) \cite{Cabibbo:1963yz,Kobayashi:1973fv} matrix characterized by four independent parameters. At present, the SM and its built-in CKM mechanism for mixing and CP violation are in good agreement with most of the experimental data. This is an impressive achievement of the SM, given the large amount of data.
The proliferation of free parameters in the SM is one of the motivations for considering New Physics (NP) which could shed some light on the flavour puzzle. On the experimental side, there is also motivation to consider NP with flavour implications; in particular, there are hints of potential
deviations from SM predictions, such as:
\newline\indent (i) A tension between the experimental value of $\AJPKs$ \cite{Aubert:2009aw,Adachi:2012et} (the time dependent CP asymmetry in $B_{d}^{0}\to J/\psi K_{S}$) and the value implied by flavour fits. In the context of the SM, $\AJPKs=\sin 2\beta$ where $\beta=\arg \left[-\Vc{td}\V{tb}\V{cd}\Vc{cb}\right]$. The experimental value of $\AJPKs$ is around $2\sigma $ lower than the value of $\sin 2\beta $ extracted indirectly from other experimental input.
\newline\indent(ii) The experimental value of the branching ratio $\BTNu$ is around $2.5\sigma$ larger than the value expected in the SM \cite{Ikado:2006un,Hara:2010dk,Aubert:2007xj,Lees:2012ju}. Notice, nevertheless, that the last analysis of the Belle collaboration \cite{Adachi:2012mm} quotes a value for $\BTNu$ much lower than previous analyses. If such a value is confirmed and persists over time, the case for NP with flavour implications would weaken.

In this paper, we examine the question whether the present data may already give some hints of New Physics (NP) and also discuss the potential for data expected in the future to give confirmation and /or provide more restrictive bounds on NP. We start our analysis by presenting a brief update of a model independent analysis (MIA) looking for the presence of NP \cite{Botella:2005fc,Botella:2006va} in view of the present experimental data. For definiteness, in the MIA, we shall make the following assumptions:
\begin{itemize}
\item[(i)] We assume that the extraction of parameters from SM tree level dominated processes is not affected by the possible presence of NP, like in: $|\V{ud}|$, $|\V{us}|$, $|\V{ub}|$, $|\V{cd}|$, $|\V{cs}|$, $|\V{cb}|$ and $\gamma =\arg \left[-\V{ud}\V{cb}\Vc{ub}\Vc{cd}\right]$.
\item[(ii)] We allow for NP to give significant contributions to \BBdmix\ and \BBsmix\ transitions.
\end{itemize}
In the MIA we shall assume that the CKM matrix $\CKM$ is a $3\times 3$ unitary matrix. Our MIA applies to a large class of theories beyond the SM, including supersymmetric extensions of the SM and some generic two Higgs doublet models. However, it should be emphasized that $3\times 3$ unitarity is an assumption which can be violated in reasonable extensions of the SM. 

The main goal of this paper is to make a study of the impact of NP on the flavour sector, in the context of a simple extension of the SM where there are naturally small violations of $3\times 3$ unitarity in $\CKM$. For definiteness, we consider an extension of the SM where an isosinglet $Q=2/3$ vector-like quark is added to the SM. The paper is organised as follow. First we present the updated MIA. In section \ref{sec:uVLQ} we define the model with a singlet up vector-like quark (\VL\!\!), then we successively present the flavour and CP results in the $bd$, $bs$ and $ds$ sectors. We devote special attention to correlations among different observables. In section \ref{sec:up} we discuss rare up decays and the mass of the new up quark. Then we present our conclusions and devote two appendices to explain additional details.

% % % % % % % % % % % % % % % % % % % % % %
\section{Model Independent Analysis\label{sec:MIA}}
% % % % % % % % % % % % % % % % % % % % % %

With the standard definitions of the rephasing invariant quantities \cite{Branco:1999fs}
\begin{align}
\alpha &=\arg \left[ -\V{td}\V{ub}\Vc{tb}\Vc{ud}\right]\,,\nonumber \\ 
\beta &=\arg \left[ -\V{cd}\V{tb}\Vc{cb}\Vc{td}\right]\,,\nonumber \\ 
\gamma  &=\arg \left[ -\V{ud}\V{cb}\Vc{ub}\Vc{cd}\right]\,,\nonumber \\ 
\beta _{s}&=\arg \left[ -\V{cb}\V{ts}\Vc{cs}\Vc{tb}\right]\,,
\label{eq:angles}
\end{align}%
$\alpha +\beta +\gamma =\pi$ by definition, independently of whether the CKM matrix $\CKM$ is $3\times 3$ unitary or not. 
It is well known that in the context of the SM, one has, to an excellent approximation:
\begin{equation}
\AJPKs=-\frac{\Gamma \left(B_{d}^{0}\to J/\psi K_{S}\right) -\Gamma \left( \bar{B}_{d}^{0}\to J/\psi K_{S}\right) }{\sin \left(\DMBd\,t\right)\, \left[ \Gamma \left( B_{d}^{0}\to J/\psi K_{S}\right) +\Gamma\left( \bar{B}_{d}^{0}\to J/\psi K_{S}\right)\right]}=\sin 2\beta\,.  \label{eq:sin2beta}
\end{equation}
In a similar way, in the SM, the time-dependent asymmetry in $B_d^0\to (\pi\pi)_{I=2}$ vs. $\bar B_d^0\to (\pi\pi)_{I=2}$, $A_{(\pi\pi)_{I=2}}$, where $(\pi\pi)_{I=2}$ denotes the strong isospin state of the pair of pions, is
\begin{equation}
A_{(\pi\pi)_{I=2}}=\sin(2\alpha)\,.\label{eq:sin2alpha}
\end{equation}
As $2\alpha=-(\beta+\gamma)\,\mod{[2\pi]}$, $A_{(\pi\pi)_{I=2}}$ can be viewed as a measurement of $\gamma$; direct measurements of $\gamma$ are achieved, anyway, in other channels \cite{Gronau:1990ra,Gronau:1991dp,Aleksan:1991nh,Atwood:1996ci,Atwood:2000ck,Giri:2003ty,Aubert:2007ii,Poluektov:2010wz}. In the SM, $\AJPKs$, $A_{(\pi\pi)_{I=2}}$ measurements and moduli of $|\V{ij}|$ are related through the unitarity relation
\begin{equation}
\left|\frac{\V{ud}\V{ub}}{\V{cd}\V{cb}}\right|\,\sin(\gamma+\beta)=\sin(\beta)\,,\label{eq:unitrel:0}
\end{equation}
which constitutes a nice consistency check of the SM flavour picture.

\begin{figure}[h]
\begin{center}
\includegraphics[height=0.2\textheight]{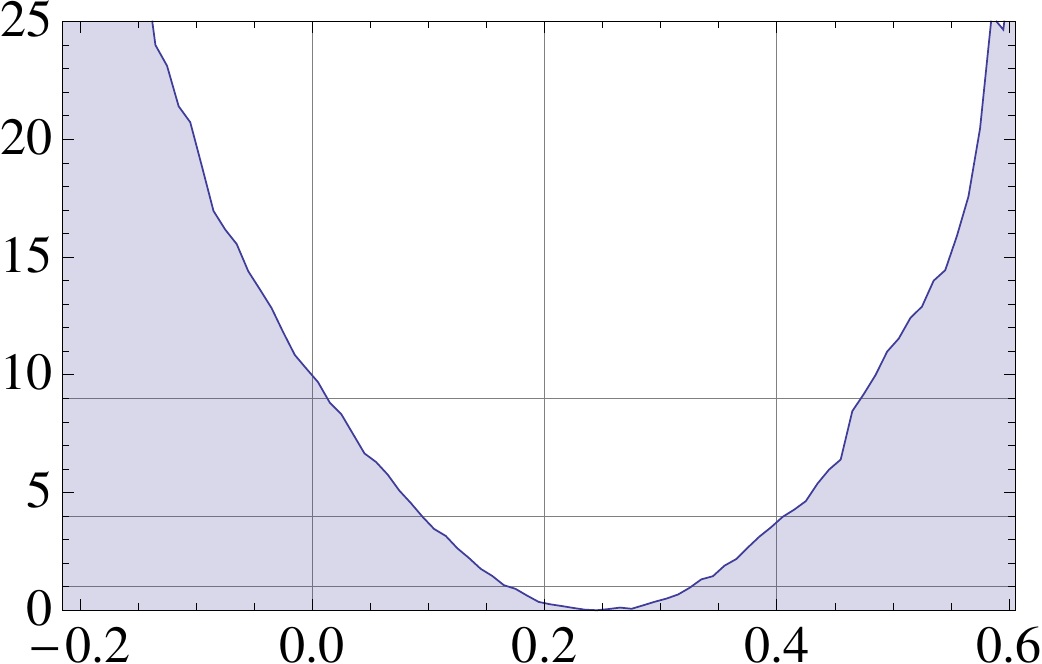}
\caption{$\Delta\chi^2$ profile of $2\phi_d$ in the MIA.}\label{fig:2phid}
\end{center}
\end{figure}

When new physics is considered, a minimal and reasonable assumption is to allow for new contributions to loop-controlled processes but not for tree level observables, while preserving $3\times 3$ unitarity of the mixing matrix. In particular, potentially modified mixings of $B$ mesons (both $B_d$ and $B_s$) can be written in the following form
\begin{equation}
\mixMBq=\left[\mixMBq\right]_{\text{SM}}\,r_q^2\,e^{-i2\phi_q}\,,\label{eq:NPmix3x3unit}
\end{equation}
where $\left[\mixMBq\right]_{\text{SM}}$ stands for the SM contribution and $\{r_q,\phi_q\}$ parameterise NP-induced deviations from SM expectations. The CP asymmetries in eq.(\ref{eq:sin2beta}) and eq.(\ref{eq:sin2alpha}) are automatically modified to
\begin{equation}
\AJPKs=\sin(2(\beta-\phi_d))\equiv \sin(2\bar{\beta})\,,\qquad A_{(\pi\pi)_{I=2}}=\sin(2(\alpha+\phi_d))=\sin(2\bar\alpha)\ .
\end{equation}
It is clear that $\gamma$, or equivalently $\alpha+\beta$, can be extracted from experiment in a model independent way, as
\begin{equation}
\pi-\gamma=\frac{1}{2}\left(\arcsin{\AJPKs}+\arcsin{A_{(\pi\pi)_{I=2}}}\right)\ .
\end{equation}
Then, using eq.(\ref{eq:unitrel:0}) with $R_u\equiv \left|\frac{\V{ud}\V{ub}}{\V{cd}\V{cb}}\right|$, one can obtain \cite{Botella:2002fr}
\begin{equation}
\tan\phi_d=\frac{R_u\,\sin(\gamma+\bar{\beta})-\sin\bar{\beta}}{\cos\bar{\beta}-R_u\,\cos(\gamma+\bar{\beta})}\,.\label{eq:tanphid}
\end{equation}
In the absence of NP, $\bar{\beta}= \beta$ and $\phi_d= 0$. Using eq.(\ref{eq:tanphid}) and taking into account experimental data, one obtains
\[
\tan\phi_d=0.11\pm 0.03\ .
\]
A complete analysis yields the result summarized in figure \ref{fig:2phid}.

%%%%%
%%%%%

The conclusion is clear: there is a tension between $R_u$, $\bar{\beta}$ and $\bar{\alpha}$, and NP in \BBdmix\ may solve or relax it. Although dominated by the SM picture, data appears to be pointing to a small but significant presence of NP in the $bd$ sector.

One can extend the analysis to the $bs$ sector \cite{Ligeti:2006pm,Ball:2006xx,Grossman:2006ce,Bona:2006sa,Botella:2006va} by considering the time-dependent asymmetry (contrary to $B_d^0\to J/\Psi\,K_S$, angular analysis to separate different CP components in the final state is required) in $B_s^0\to J/\Psi\Phi$ vs. $\bar B_s^0\to J/\Psi\Phi$, $\AJPP$, which in the SM is
\begin{equation}
\AJPP=\sin(2\beta_s)\,,
\end{equation}
expected to be small, since $3\times 3$ unitarity, in the CKM framework, forces $\beta_s= 0.187\pm 0.006$. Allowing for the modification in eq.(\ref{eq:NPmix3x3unit}) gives
\begin{equation}
\AJPP=\sin(2(\beta_s+\phi_s))\, .
\end{equation}
It is, therefore, the LHCb measurement of $\AJPP$ --- and related channels --- that dominates the extraction of $\phi_s$ \cite{LHCb:2011aa,LHCb:2011ab}.
There is ample room for NP in the $bs$ sector --- $\phi_s$ can be two or three times larger than $\beta_s$ ---. However, the current situation, dominated by experimental uncertainty while being close to SM expectations does not require the presence of NP.

% % % % % % % % % % % % % % % % % % % % % %
\section{Up vector-like singlet quark model\label{sec:uVLQ}} %{Model with one Q=2/3 vector-like singlet quark (sUVL): Introduction}
% % % % % % % % % % % % % % % % % % % % % %
%
\indent For definiteness, we consider an extension of the SM where one isosinglet vector-like quark $T$ with charge $Q=2/3$ is added to the spectrum of the SM \cite{Langacker:1988ur,delAguila:1982fs,delAguila:1987nn,Cheng:1991rr,delAguila:1985mk,delAguila:1985ne,Branco:1986my,Buchmuller:1988et,Nir:1990yq,Nardi:1991rg,Silverman:1991fi,Branco:1992uy,Branco:1992wr,Branco:1995us,Barger:1995dd,delAguila:1997vn,Barenboim:1997pf,Barenboim:1997qx,Kakebe:1997bj,Barenboim:2000zz,Higuchi:2000rh,Barenboim:2001fd,AguilarSaavedra:2002kr,AguilarSaavedra:2004mt,Botella:2008qm,Higuchi:2009dp,Frampton:1999xi}. The $3\times 3$ mixing matrix connecting standard quarks is no longer unitary, but a submatrix of a $4\times 4$ unitary matrix $U$. Without loss of generality, one can choose to work in the weak basis (WB) where the down quark mass matrix is diagonal and real. In this basis, $U$ is a $4\times 4$ unitary matrix which enters the bidiagonalization of the $4\times 4$ mass matrix $\mathcal{M}$ of $Q=2/3$ quarks:
\begin{equation}
U^{\dagger }\mathcal{MM}^{\dagger }U=\text{diag.}\left(
m_{u}^{2},m_{c}^{2},m_{t}^{2},{\mT}^{2}\right)\,.  \label{eq:UPdiagonalization}
\end{equation}
The charged and neutral current interactions have the form
\begin{equation}
\mathscr{L}_{W}=-\frac{g}{\sqrt{2}}\mathbf{\bar{u}}_{L}\,\gamma ^{\mu}\,V\,\mathbf{d}_{L}\,W_{\mu }+\text{h.c.}\,,  \label{eq:ChargedCurrents:00}
\end{equation}
\begin{equation}
\mathscr{L}_{Z}=-\frac{g}{2\cos\theta_{W}}\left[ \mathbf{\bar{u}}_{L}\,\gamma^{\mu }\left(VV^{\dagger}\right)\,\mathbf{u}_{L}-\mathbf{\bar{d}}_{L}\,\gamma^{\mu}\,\mathbf{d}_{L}-2\sin ^{2}\theta_{W}\,J_{em}^{\mu}\right]\, Z_{\mu}\,,  \label{eq:NeutralCurrents:00}
\end{equation}%
where $\mathbf{d}\equiv \left(d,s,b\right)$, $\mathbf{u}\equiv\left(u,c,t,T\right)$ and $V$ is a $4\times 3$ submatrix of the matrix $U$:
\begin{equation}
U=\left( 
\begin{array}{cccc}
V_{ud} & V_{us} & V_{ub} & U_{u4} \\ 
V_{cd} & V_{cs} & V_{cb} & U_{c4} \\ 
V_{td} & V_{ts} & V_{tb} & U_{t4} \\ 
V_{Td} & V_{Ts} & V_{Tb} & U_{T4}
\end{array}
\right)\ .  \label{eq:mixmatrixU:00}
\end{equation}%
It is clear that the submatrix $V_{(3\times 3)}$, i.e. the upper left $3\times 3$ block within $U$, is not a unitary matrix, since $V_{(3\times 3)}^{\phantom{\dagger}}\, V_{(3\times 3)}^{\dagger}\neq \mathbf{1}_{(3\times 3)}$.
However, these deviations of unitarity of the ``would-be standard'' mixing matrix are naturally suppressed by the ratio $m^{2}/m_{T}^{2}$, where $m$ denotes generically the standard quark masses \cite{Langacker:1988ur,delAguila:1982fs,delAguila:1987nn,Cheng:1991rr}. These deviations from unitarity lead to flavour changing neutral currents (FCNC) which are present just in the up sector and controlled by 
\begin{equation}
\left( VV^{\dagger }\right)_{ij}=\delta _{ij}-U_{i4}^{\phantom{\ast}}U_{j4}^{\ast}\ .
\label{eq:NeutralCurrents:01}
\end{equation}
These FCNC are thus suppressed by the ratio $m^2/\mT^2$. This natural suppression of FCNC is crucial in order to make the model plausible, since FCNC mediate dangerous tree-level processes leading for example to \DDmix\ mixing though $U_{u4}^{\phantom{\ast}}U_{c4}^{\ast}$.

%\textbf{Comparison with the 4-generations Standard model}
% % % % % % % % % % % % % % % % % % % % % %
\subsection{Comparison with the 4-generations Standard model\label{ssec:SM4}} 
% % % % % % % % % % % % % % % % % % % % % %
At this stage, it is instructive to compare the \VL model with the four generations Standard Model (SM4). The differences between these two minimal
extensions of the SM can be summarized as follows:

\begin{enumerate}
\item Both the \VL model and the SM4 involve deviations of $3\times 3$ unitarity in $\CKM$. In the \VL model deviations of unitarity are naturally suppressed by the ratio $m^{2}/\mT^{2}$ and lead to naturally small $Z$-mediated FCNC in the up sector. In the SM4 there are no tree level FCNC due to an exact GIM mechanism \cite{Glashow:1970gm}.
\item The dominant Higgs boson production mechanism at the LHC is the gluon-gluon fusion process which is dominated by a heavy quark loop \cite{Wilczek:1977zn,Georgi:1977gs}. With the two new heavy sequential quarks introduced in the SM4, there is the potential of having an increase in the production cross section by a factor of $(1+2)^{2}=9$. Even if smaller, this enhancement has very important consequences \cite{Gonzalez:2011he,Passarino:2011kv,Djouadi:2012ae,Eberhardt:2012ck}. In the \VL model there is no enhancement, since the quark mass
terms of the vector-like quark do not arise from electroweak symmetry breaking.
\end{enumerate}

\clearpage
% % % % % % % % % % % % % % % % % % % % % % % % %
%\section{The $\boldsymbol{bd}$ sector \label{sec:bd}}
\section{The ${bd}$ sector \label{sec:bd}}
\begin{figure}[H]
\begin{center}
\includegraphics[width=0.42\textwidth]{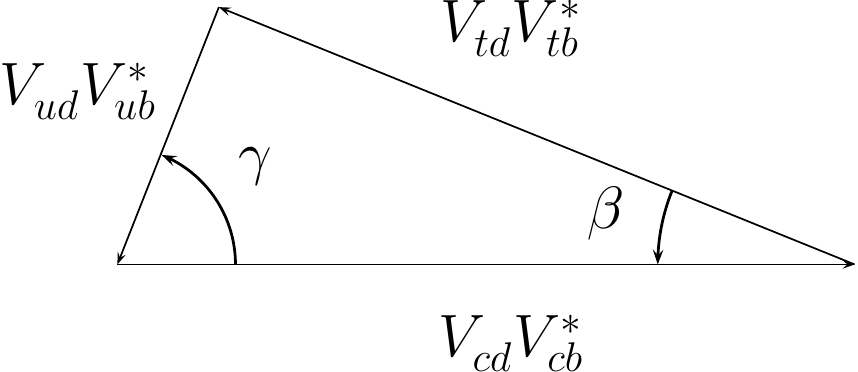}
\includegraphics[width=0.42\textwidth]{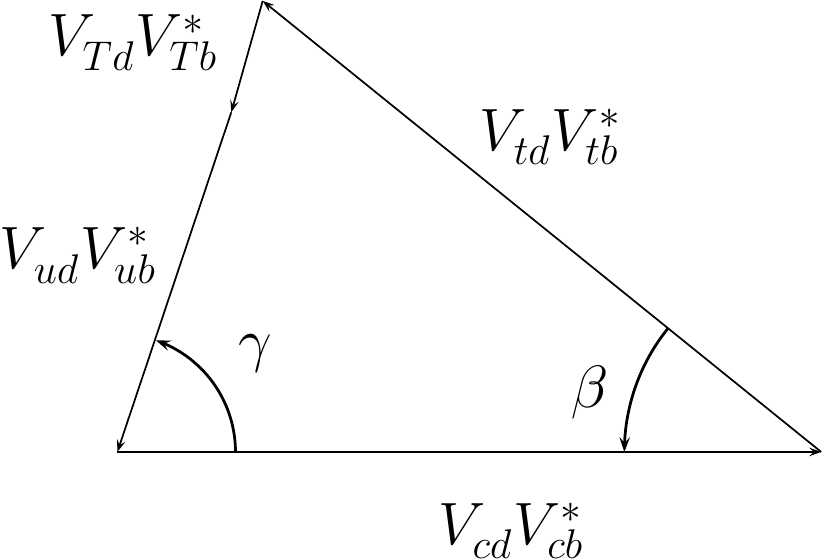}
\caption{Unitarity quadrangle vs. unitarity triangle.\label{fig:unitarity}}
\end{center}
\end{figure}

\begin{figure}[H]
\begin{center}
\includegraphics[width=0.6\textwidth]{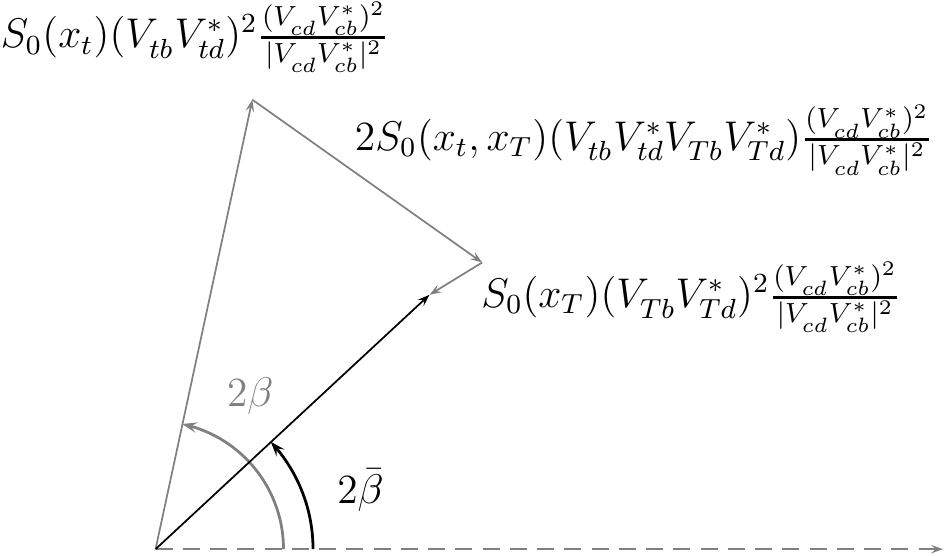}
\caption{Pictorial representation of the different contributions to $\mixMBd$ in the \VL model.}\label{fig:4x4mix}
\end{center}
\end{figure}

\begin{figure}[h]
\begin{center}
\subfigure[SM case.\label{fig:BTauNu:vs:AJPKs:SM}]{\includegraphics[height=0.28\textheight]{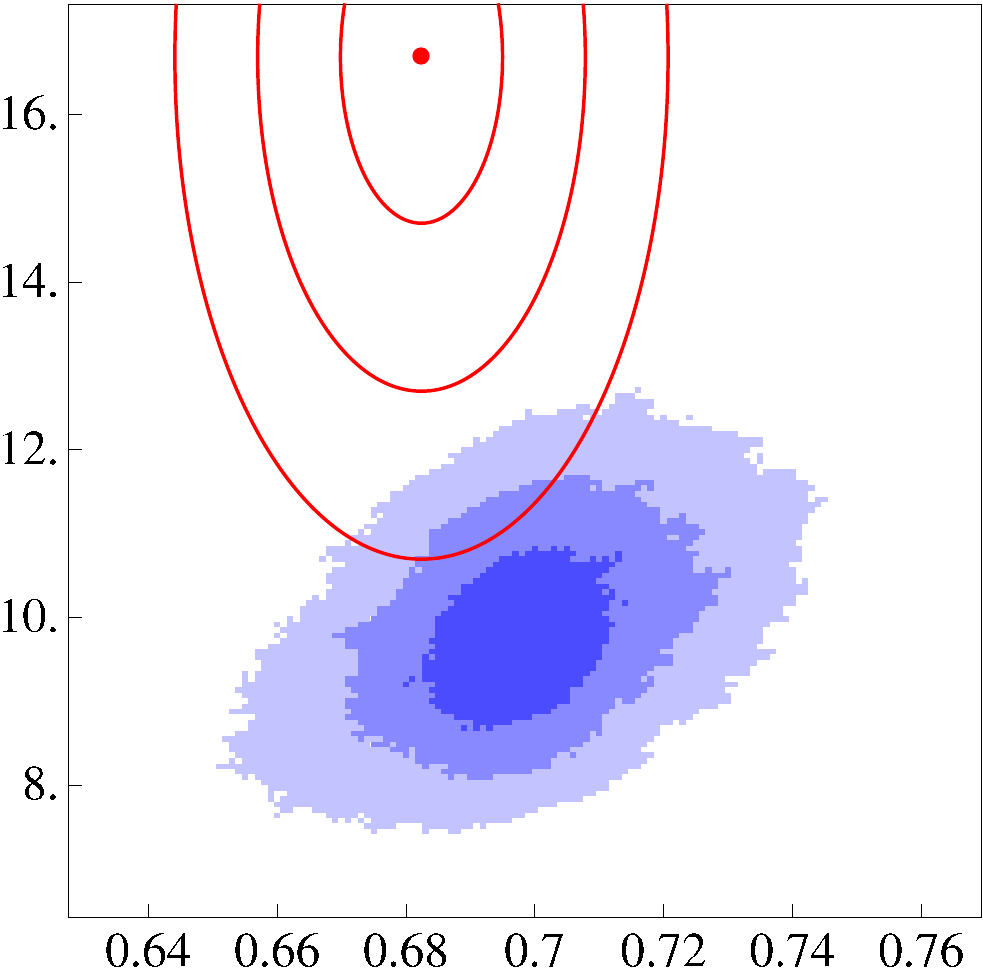}}\qquad
\subfigure[\VL model case.\label{fig:BTauNu:vs:AJPKs:VL}]{\includegraphics[height=0.28\textheight]{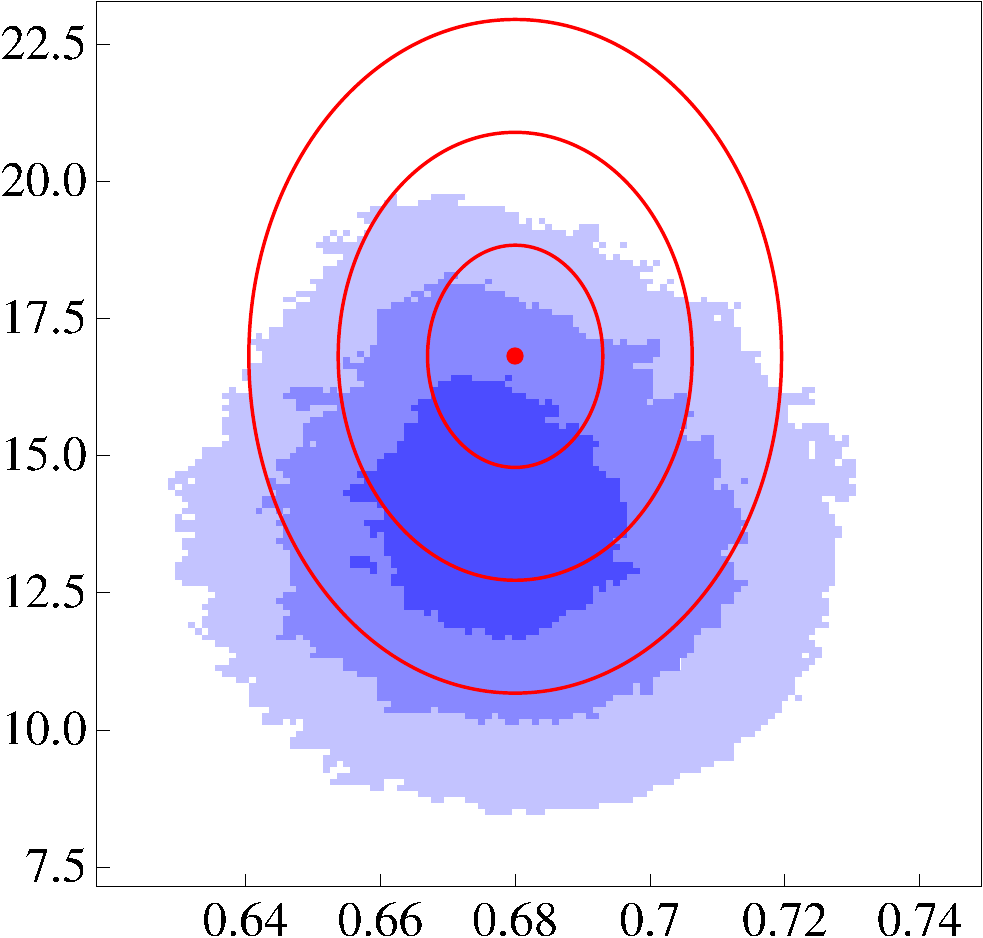}}
\caption{$\BTNu\times 10^{5}$ vs. $\AJPKs$, 68\%, 95\% and 99\% CL regions (darker to lighter); the ellipses show the 68\%, 95\% and 99\% CL regions corresponding to the measurement.\label{fig:BTauNu:vs:AJPKs:both}}
\end{center}
\end{figure}

\begin{figure}[h]
\begin{center}
\subfigure[$|V_{ub}|\times 10^3$.\label{fig:Vub:VL}]{\includegraphics[height=0.2\textheight]{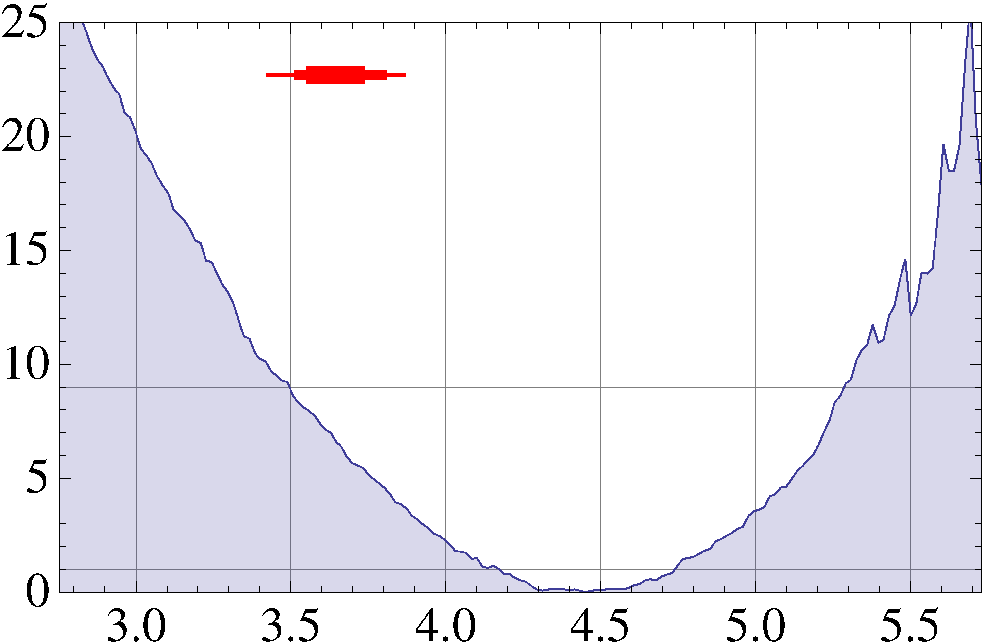}}\qquad
\subfigure[$\AJPKs$.\label{fig:AJPKs:VL}]{\includegraphics[height=0.2\textheight]{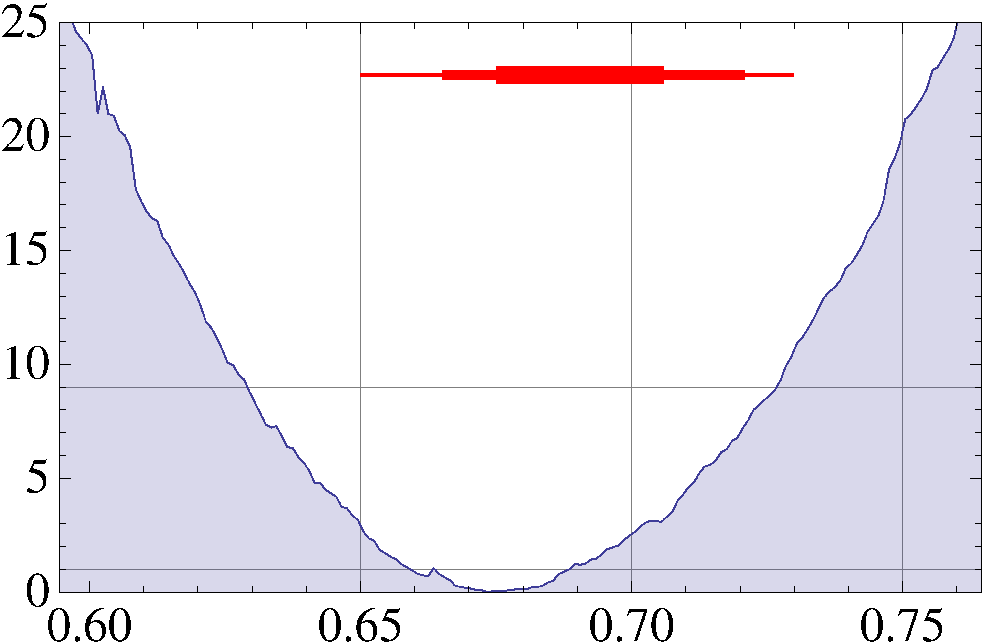}}
\caption{$\Delta\chi^2$ profiles of $|\Vfig{ub}|$ and $\AJPKs$ in the \VL model. Red bars (thicker to thinner) show the $1$, $2$ and $3$ $\sigma$ ranges corresponding to a SM fit. The experimental values are $|\Vfig{ub}|=(3.89\pm 0.44)\times 10^{-3}$ and $\AJPKs=0.68\pm 0.02$.}\label{fig:Vub:AJPKs:VL}
\end{center}
\end{figure}

% % % % % % % % % % % % % % % % % % % % % % % % %
As a starter, let us analyse how the \VL model is able to mitigate the tensions present in the $bd$ sector of the SM. Figure \ref{fig:unitarity} displays the usual unitarity triangle (SM) together with a realistic quadrangle corresponding to the \VL model. It is important to stress that, in order to reduce the tension, we need an increase of $|\V{ub}|$ to enlarge $\BTNu$. But, in the SM, this comes hand in hand with an enlarged $\beta$, thus generating a conflict, a tension, with the experimental value \cite{Bona:2009cj,Lunghi:2010gv}. This tension can be solved by the necessary NP in the mixing, which, in this model, is fixed by the enlarged mixing matrix and the mass of the heavy quark. Focusing on the $bd$ sector, this means that it is fixed by the shape of the $bd$ unitarity quadrangle. So, in practice, one has more freedom by increasing $|\V{ub}|$, increasing $\beta$ and adding the fourth side $|\V{Td}\Vc{Tb}|$, necessary to match the potential modifications in $|\V{ub}|$ and $\beta$.
%%%%%%%
%%%%%%%
A particular case of how the NP in the mixing transforms $\beta $ into $\bar{\beta}$, corresponding to the previous quadrangle, is shown in the figure \ref{fig:4x4mix}. Notice how the SM term --- the one to the left, oriented with respect to the horizontal $\left(\V{cd}\Vc{cb}\right)^{2}$ ---, corresponding to the box diagram with internal $t$ quarks, receives additional contributions from the box diagrams with both $T$ and $t$ and only with $T$, respectively. As one can see, starting from a large $2\beta $ one ends up with a smaller $2\bar{\beta}$. The different $S_0$ functions are the well-known Inami-Lim functions \cite{Inami:1980fz}.
In order to show that the \VL model certainly solves the tensions in the $bd$ sector we first present together the results of the SM-CKM and \VL fits in the $(\BTNu,\AJPKs)$ plane in figure \ref{fig:BTauNu:vs:AJPKs:both}. The experimental input used, together with the methodology and further details, are explained in the appendices. Let us just clarify here that, in figure \ref{fig:BTauNu:vs:AJPKs:both} (as in many other figures along the present paper), are represented three (blue) regions corresponding to 68\%, 95\% and 99\% CL (darker to lighter). The ellipses represent the same CL regions for experimental data.
%%%%%%%
%%%%%%%
Figure \ref{fig:BTauNu:vs:AJPKs:SM} clearly shows the tensions among these observables in the SM case. In figure \ref{fig:BTauNu:vs:AJPKs:VL} it is easy to appreciate how the \VL model solves the tensions in the $bd$ sector.
%%%%%%%
%%%%%%%
%
Of course, in order to obtain in this model the experimental value of $\BTNu$, the only possibility is through an increase of $|\V{ub}|$, as can be seen in figure \ref{fig:Vub:VL}; this is accomplished without changing $\AJPKs$, as shown in figure \ref{fig:AJPKs:VL}.
%%%%%
%%%%%

\begin{figure}[H]
\begin{center}
\includegraphics[height=0.28\textheight]{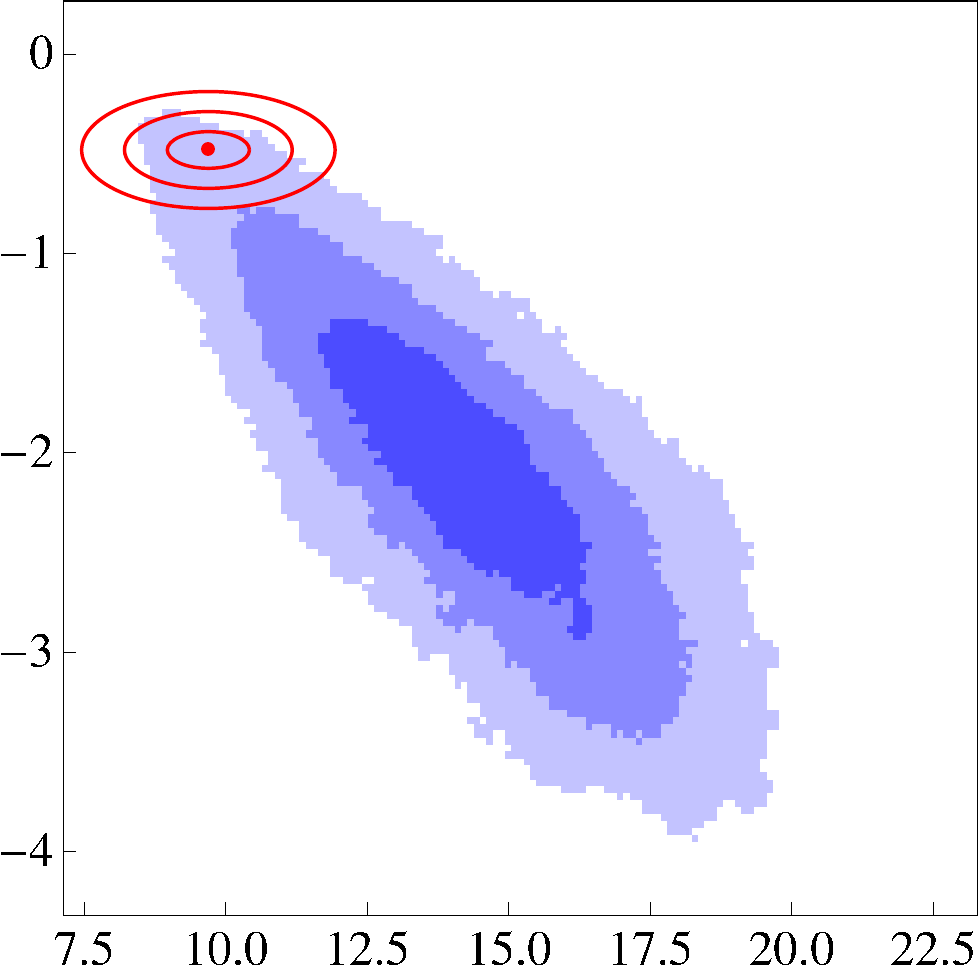}
\caption{$\asld\times 10^3$ vs. $\BTNu \times 10^5$, 68\%, 95\% and 99\% CL regions (darker to lighter); the ellipses show the 68\%, 95\% and 99\% CL regions in the SM fit. The experimental values are $\asld=(-3.0\pm 7.8)\times 10^{-3}$ and $\BTNu=(16.8\pm 3.1)\times 10^{-5}$.}\label{fig:ASLd:vs:BTauNu:VL}
\end{center}
\end{figure}

A first consequence is that this can be done by replacing the $bd$ triangle by a $bd$ quadrangle whose fourth side is given by $|\V{Td}\V{Tb}|$.
%%%%%%%
\begin{figure}[H]
\begin{center}
\includegraphics[height=0.28\textheight]{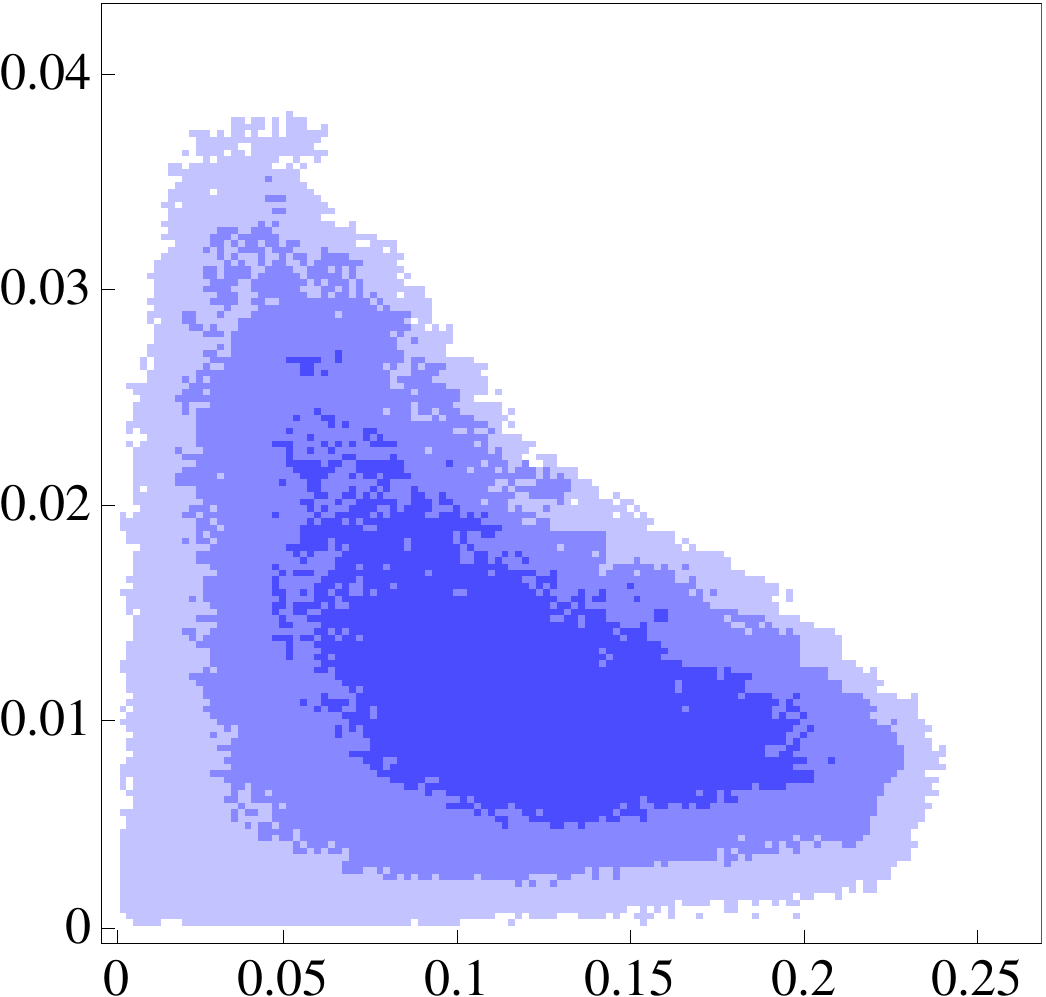}
\caption{$|\Vfig{Td}|$ vs.$|\Vfig{Tb}|$, 68\%, 95\% and 99\% CL regions (darker to lighter) in the \VL model.}\label{fig:VTd:vs:VTb}
\end{center}
\end{figure}
%%%%%%%
Figure \ref{fig:VTd:vs:VTb} shows that a typical value of the fourth side of the quadrangle is $|\V{Td}\V{Tb}| \sim \lambda^{4}$. This fourth side allows to enlarge $|\V{ub}|$ and to modify $\beta$ in such a way that the final value of $\bar{\beta}$ yields $\AJPKs$ in perfect agreement with the experimental value.
In addition to the dispersive part of the mixing, $\mixMBd$, the absorptive part $\mixGBd$ deserves attention, since it controls the semileptonic asymmetry $\asld$, $\asld=\im{\mixGBd/\mixMBd}$.
Following \cite{Botella:2006va}, it is convenient to write down the semileptonic asymmetry for $B_{d}^{0}$ in terms of physical observables\footnote{Where $K_{d}\equiv 10^{-4}\frac{G_F^2M_W^2}{12\pi^2}m_{B_d}f_{B_d}^2B_{B_d}\eta_B S_0(x_t)$, $a$, $b$ and $c$ are constants; $a$, $b$ and $c$ come from the calculation of the absorptive part of the mixing with intermediate up and charm quarks, including hadronic matrix elements, according to references \cite{Beneke:1998sy,Beneke:2003az,Ciuchini:2003ww,Lenz:2006hd}.},
\begin{multline}
\asld =\frac{K_{d}}{\DMBd}\left[\left( b+c-a\right)\left\vert \V{ud}\V{ub}\right\vert ^{2}\sin \left( 2\bar{\alpha }\right) +\right.   \\
\left. +\left( 2c-a\right) \left\vert \V{ud}\V{ub}\V{cd}\V{cb}\right\vert
\sin \left( 2\bar{\beta }+\gamma \right) -c\left\vert \V{cd}\V{cb}\right\vert ^{2}\AJPKs\right]\,.\label{eq:SL:asym:Bd}
\end{multline}%
It has to be stressed here that we have included the NP that this model introduces in $\mixMBd$, but the model does not introduce any new contribution in $\mixGBd$. Of course, one must be careful in not using any expression where $3\times 3$ unitarity has been used \cite{Beneke:1998sy,Beneke:2003az,Ciuchini:2003ww,Lenz:2006hd,Lenz:2012mb}. The first thing to note in eq.(\ref{eq:SL:asym:Bd}) is that the quadratic term in $|\V{ub}|$ has a coefficient $\sin\left( 2\bar{\alpha}\right) =0.00\pm 0.15$ so, at the leading order, $\asld$ is linear in $|\V{ub}|$. It was also noted in \cite{Botella:2006va} that there is a strong cancellation among the second and third terms so one can guess that a 10-20\% enhancement in $|\V{ub}|$ can be easily translated into an enhancement factor of five or more in $\asld$. This can be easily in seen in figure \ref{fig:ASLd:vs:BTauNu:VL}.

%%%%%%%
%%%%%%%

\begin{figure}[H]
\begin{center}
\subfigure[SM case.\label{fig:fig:beta:SM}]{\includegraphics[height=0.2\textheight]{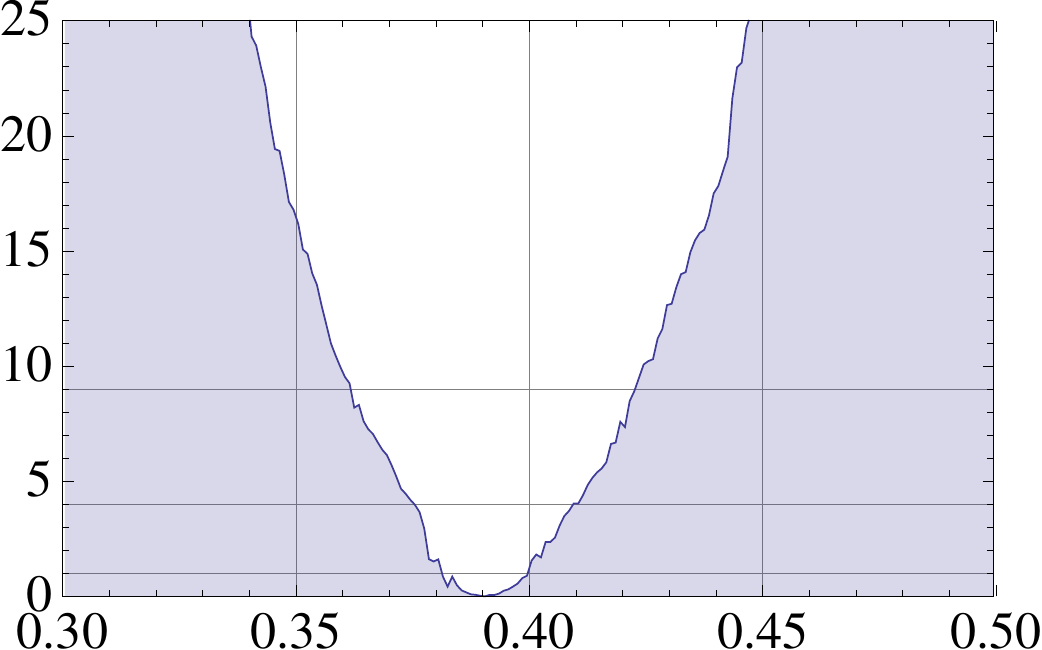}}\quad
\subfigure[\VL case.\label{fig:fig:beta:VL}]{\includegraphics[height=0.2\textheight]{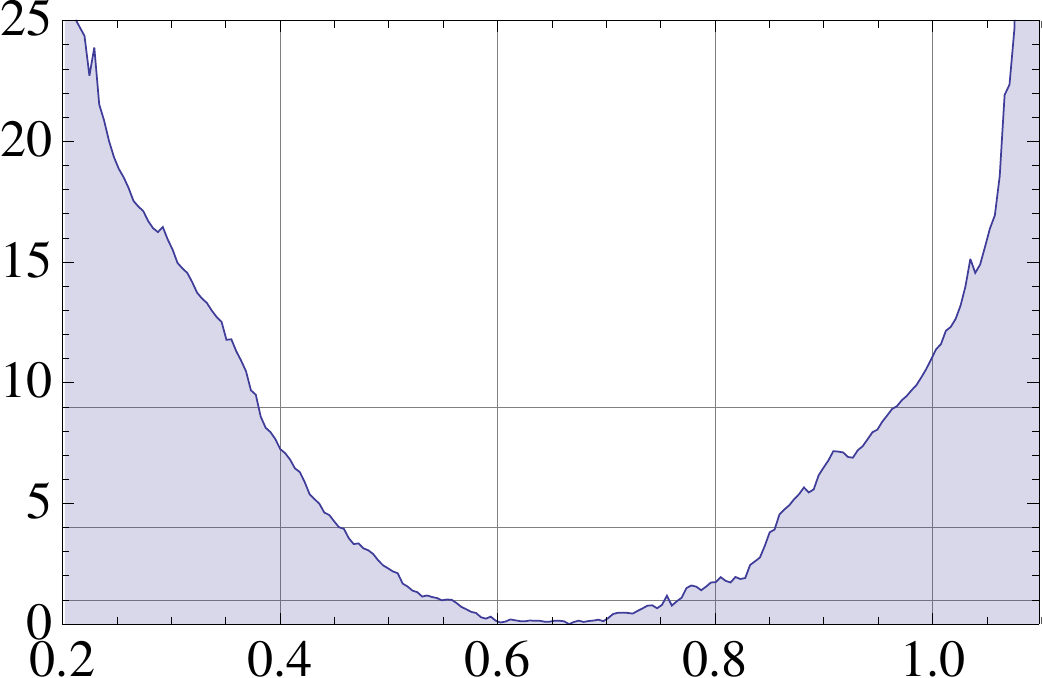}}
\caption{$\Delta\chi^2$ profile of the phase $\beta$.}\label{fig:beta:SM:VL}
\end{center}
\end{figure}

\begin{figure}[H]
\begin{center}
\subfigure[SM case.\label{fig:Bdmumu:SM}]{\includegraphics[height=0.2\textheight]{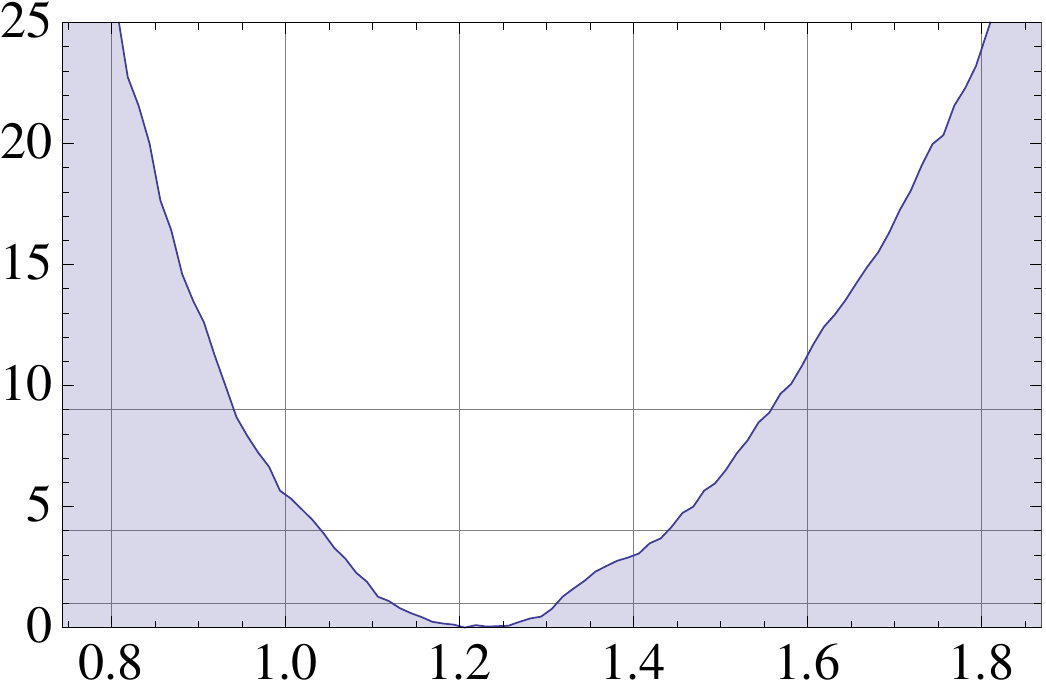}}\qquad
\subfigure[\VL case.\label{fig:Bdmumu:VL}]{\includegraphics[height=0.2\textheight]{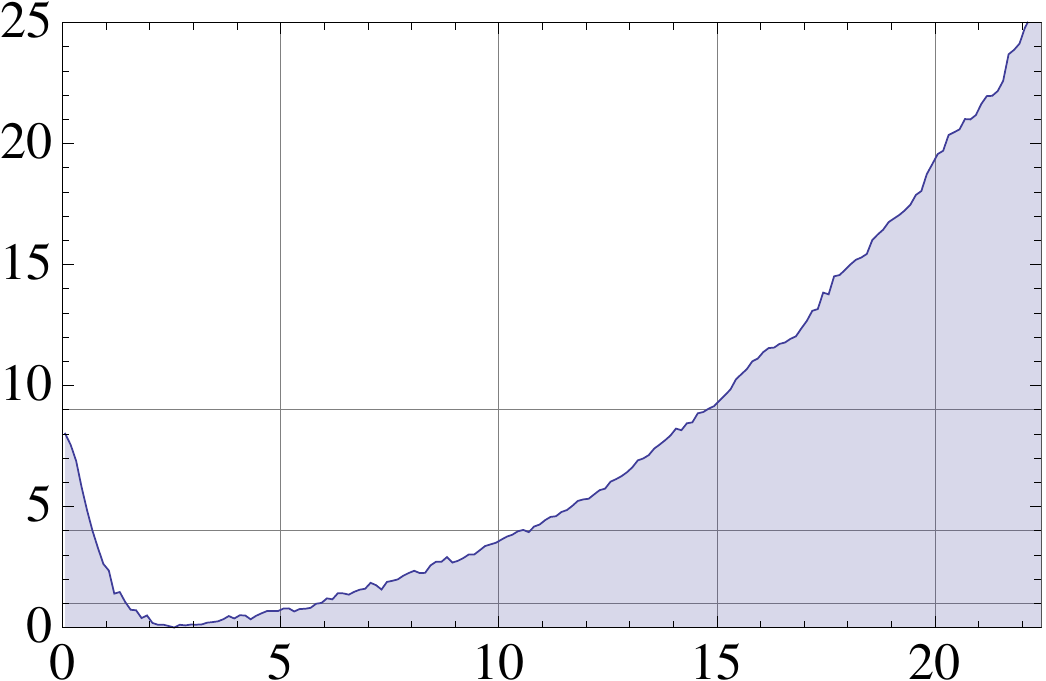}}
\caption{$\Delta\chi^2$ profile of $\Bdmm\times 10^{10}$. The experimental bound is $\Bdmm<10.5\times 10^{10}$ at 90\% CL.\label{fig:Bdmumu:SM:VL}}
\end{center}
\end{figure}

\begin{figure}[H]
\begin{center}
\subfigure[$\Bdmm\times 10^{10}$  vs. $\BTNu\times 10^5$\label{fig:Bdmm:vs:BTauNu:VL}]%
{\includegraphics[height=0.28\textheight]{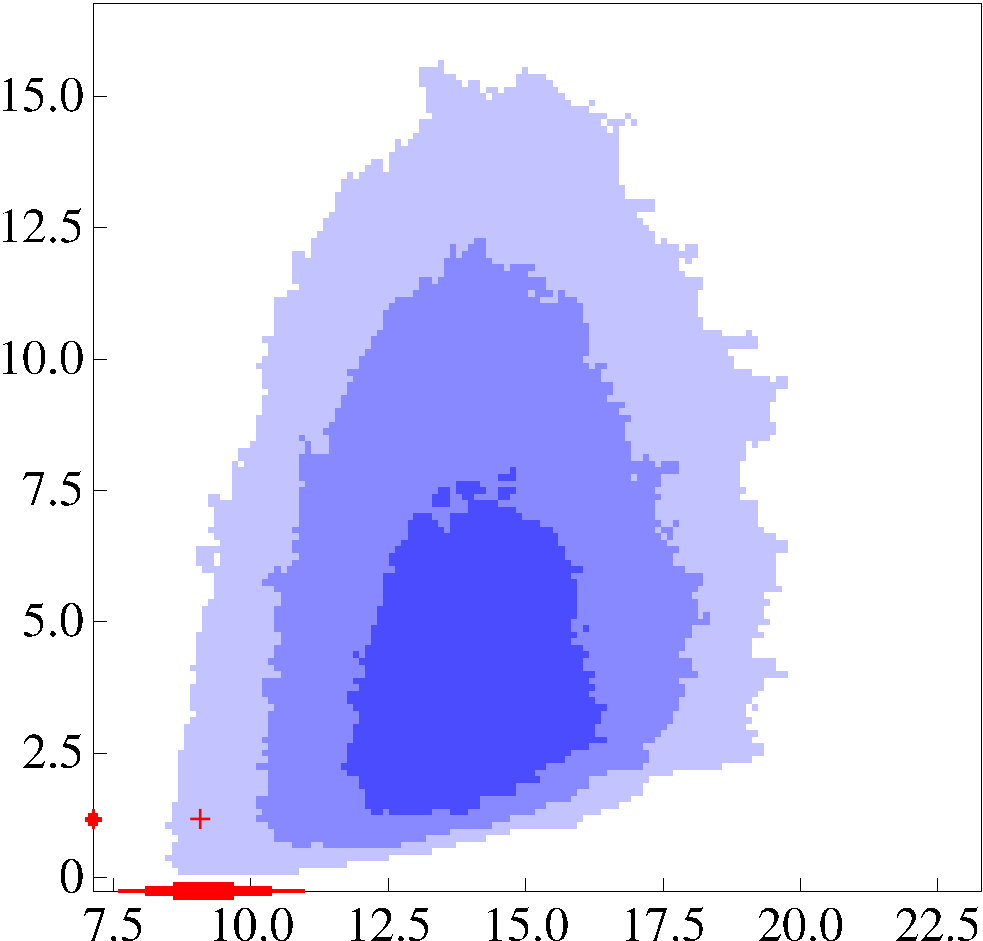}}\qquad
\subfigure[$\asld\times 10^3$ vs. $\Bdmm\times 10^{10}$\label{fig:fig:Bdmm:vs:ASLd:VL}]%
{\includegraphics[height=0.28\textheight]{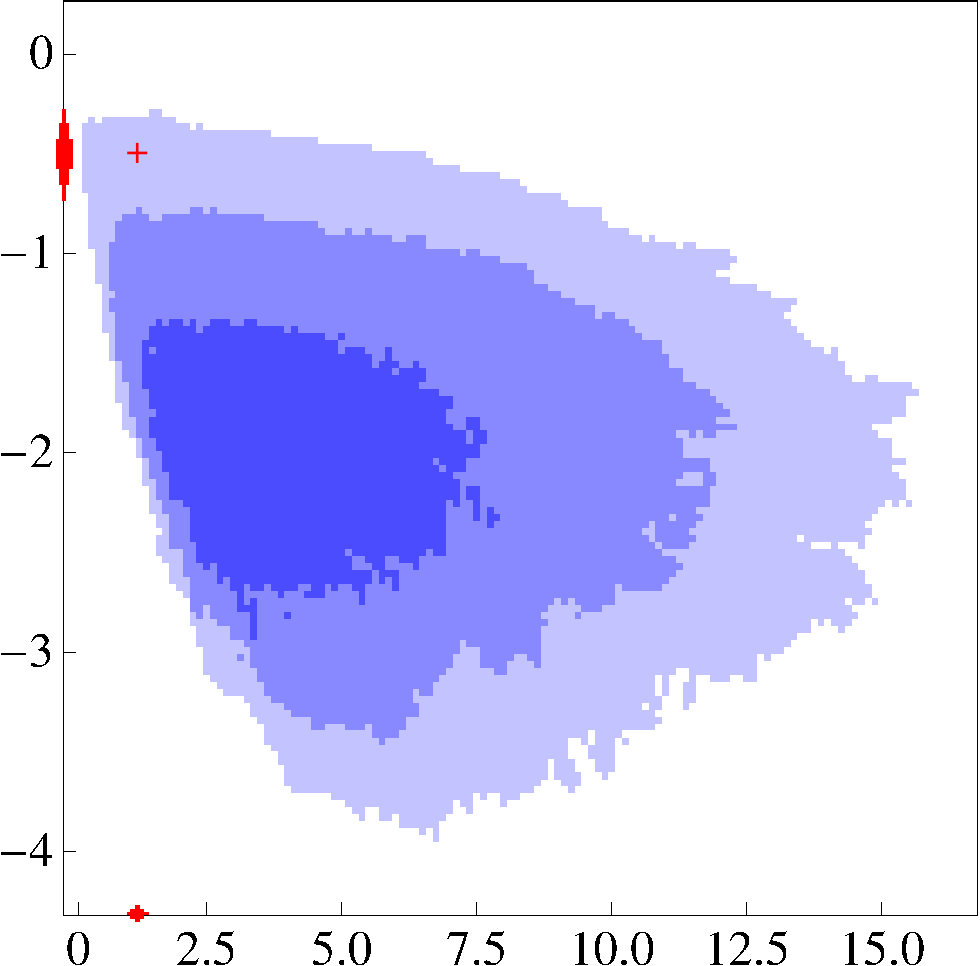}}
\caption{Correlations in the \VL model, 68\%, 95\% and 99\% CL regions (darker to lighter) are shown. Red bars (thicker to thinner) show the $1$, $2$ and $3$ $\sigma$ ranges corresponding to a SM fit. The experimental inputs are the bound $\Bdmm<10.5\times 10^{10}$ at 90\% CL and the value $\BTNu=(16.8\pm 3.1)\times 10^{-5}$.}\label{fig:Bdmm:vs:var:VL}
\end{center}
\end{figure}

This opens the door to an important contribution to the dimuon asymmetry observed by the D0 collaboration \cite{Abazov:2011yk}. We will come back to this point in the sequel. Nevertheless, it has to be pointed out that this is a common feature of any model where one enlarges $|\V{ub}|$ and introduces NP in the \BBdmix\ mixing in order to compensate the potential mismatch of $\AJPKs$ coming from a $\beta$ different from $\bar{\beta}$. 
To have a clear picture of the most important changes in the unitarity triangle --- now a quadrangle --- we plot $\beta$ in the SM and in the \VL model fits in figures \ref{fig:fig:beta:SM} and \ref{fig:fig:beta:VL} respectively.
%%%%%%
%%%%%
Another important consequence of having a relatively large value of $|\V{Td}\V{Tb}|$ is an important impact in $b\to d$ transitions like in the process $B_{d}\to \mu^{+}\mu^{-}$. We plot, in figure \ref{fig:Bdmumu:SM:VL}, the likelihood profiles of $\Bdmm$ both in the SM and in the \VL model.
%%%%%%
%%%%%
It is clear that a large enhancement --- a factor of five to ten --- still is possible in this process. It is remarkable that the last LHCb upper bound is the most important constraint on this process \cite{LHCb:2011ac,Aaij:2012ac,Chatrchyan:2011kr,Chatrchyan:2012rg,Aad:2012pn}. In the future $B^{+}\to \pi ^{+}\mu ^{+}\mu ^{-}$, for which LHCb recently announced a preliminary result, will also have a role to play. From the correlation among $\Bdmm$ and $\BTNu$, shown in figure \ref{fig:Bdmm:vs:BTauNu:VL}, it is clear that keeping $\BTNu$ around its actual experimental central value implies that $\Bdmm$ has to be seen at a rate larger than the SM value.

A similar correlation can be observed among large negative $\asld$ values and having $\Bdmm$ larger than the SM value. This can be seen in figure \ref{fig:fig:Bdmm:vs:ASLd:VL}.
%%%%%%
%%%%%

\clearpage
% % % % % % % % % % %
%\section{The $\boldsymbol{bs}$ sector\label{sec:bs}} %{The $\boldsymbol{B_{s}^{0}}$ sector of the sUVL}
\section{The ${bs}$ sector\label{sec:bs}} %{The $\boldsymbol{B_{s}^{0}}$ sector of the sUVL}
\begin{figure}[H]
\begin{center}
\subfigure[$\asls\times 10^4$ vs. $\AJPP$.\label{fig:ASLs:vs:AJPsiPhi:VL}]{\includegraphics[height=0.28\textheight]{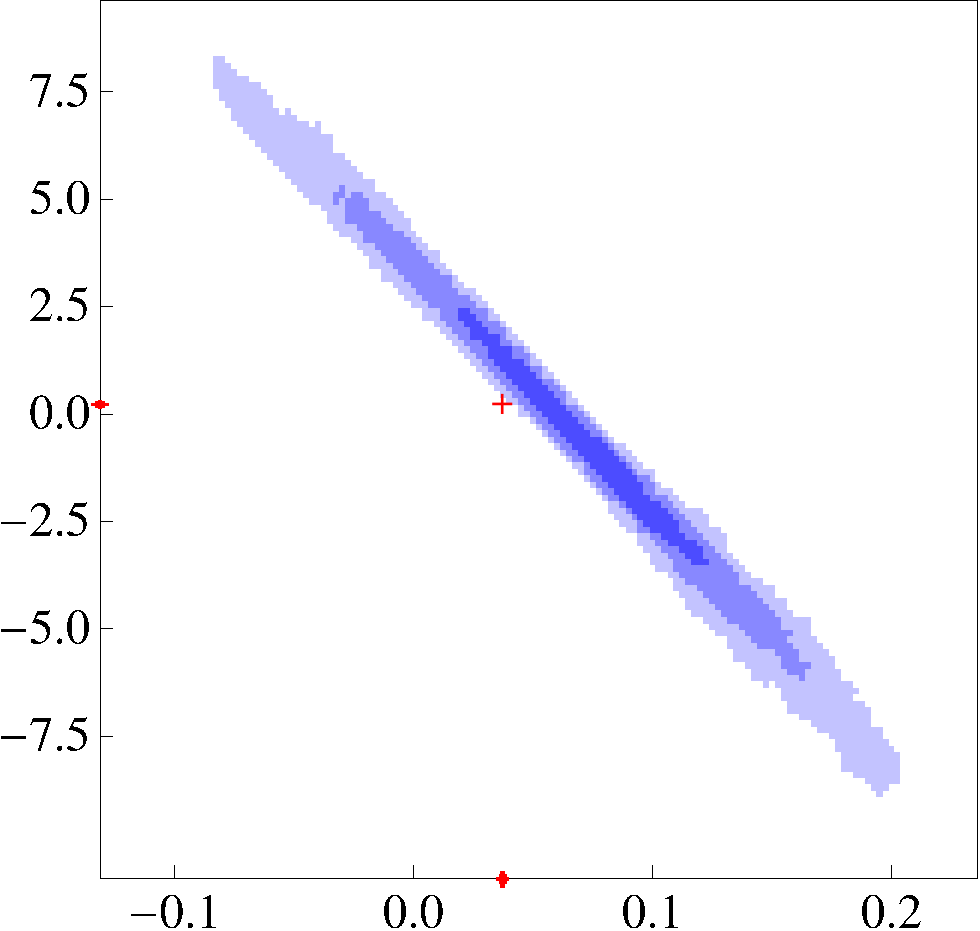}}\qquad
\subfigure[$\beta_s$ vs. $\AJPP$.\label{fig:betaS:vs:AJPsiPhi:VL}]{\includegraphics[height=0.28\textheight]{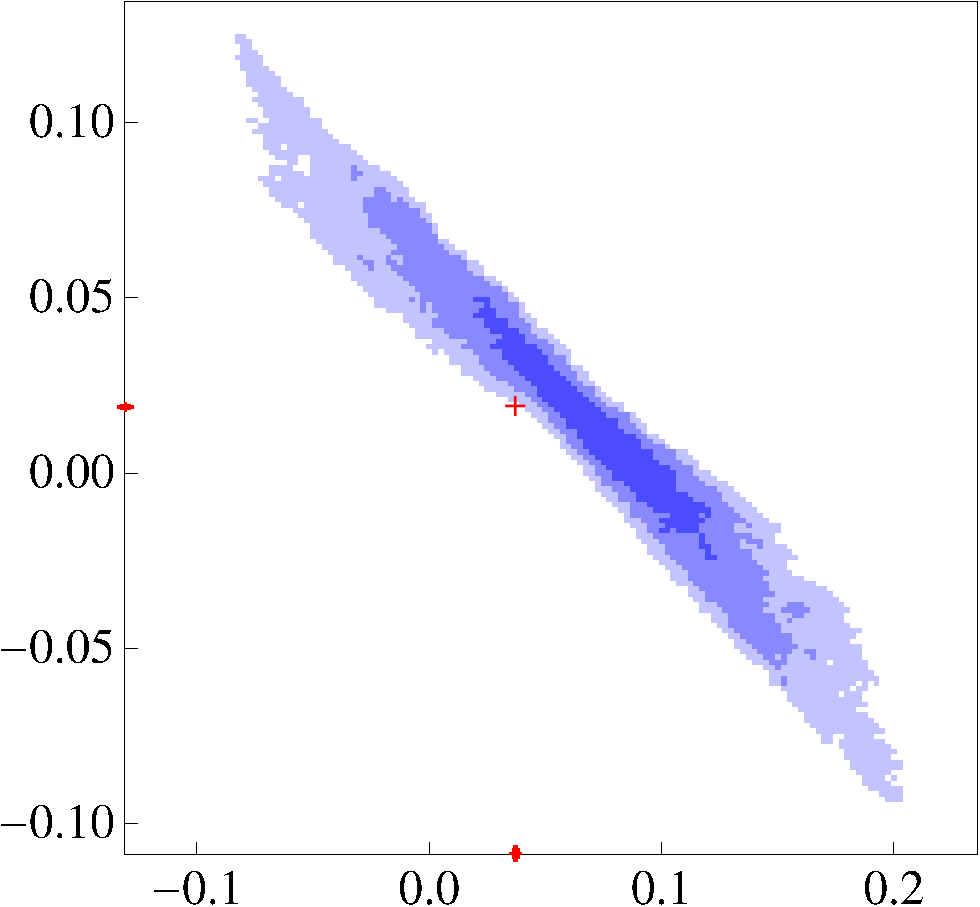}}
\caption{Correlations in the \VL model involving $\AJPP$; 68\%, 95\% and 99\% CL regions (darker to lighter) are shown. Red bars (thicker to thinner) show the $1$, $2$ and $3$ $\sigma$ ranges corresponding to a SM fit. The experimental values are $\asls=(-1.7\pm 9.1)\times 10^{-3}$ and $\AJPP=0.002\pm 0.0873$.}\label{fig:fig:var:vs:AJPsiPhi:VL}
\end{center}
\end{figure}

% % % % % % % % % % %
Last year, LHCb announced, at the Lepton-Photon Conference, its first measurement of the CP asymmetry $\AJPP$ to final CP eigenstates in the decay $B_{s}^{0}\to J/\psi \Phi $ \cite{LHCb:2011aa,LHCb:2011ab}. The associated value of the CP violating phase, $2\beta_{s}$ in the SM, was restricted to be small: this was qualified by some speaker as the Lepton-Photon drama. This designation can be understood looking at the correlation plot \ref{fig:ASLs:vs:AJPsiPhi:VL}, where we can see the 68\%, 95\% and 99\% CL regions in the plane $\{\AJPP,\asls\}$, with $\asls$ the semileptonic asymmetry in $B_{s}^{0}$ decays.
%%%%%%

%%%%%%
\vspace{-0.5cm}
%%%%%
If the large D0 dimuon asymmetry $\aslb$ had to be explained mainly through a large $\asls$ coming from NP in the $bs$ sector, this would imply also a large $\AJPP$, in conflict with the LHCb result. The origin of the name ``Drama'' comes from the fact that previous Tevatron results \cite{Abazov:2011ry,CDF:2011af} were compatible with large values of $\AJPP$ \cite{Botella:2006va}. As one can see in figure \ref{fig:ASLs:vs:AJPsiPhi:VL}, once the LHCb results on $\AJPP$ are included, $\asls$ is constrained to be in the $10^{-4}$ range.

\begin{figure}[H]
\begin{center}
\includegraphics[height=0.28\textheight]{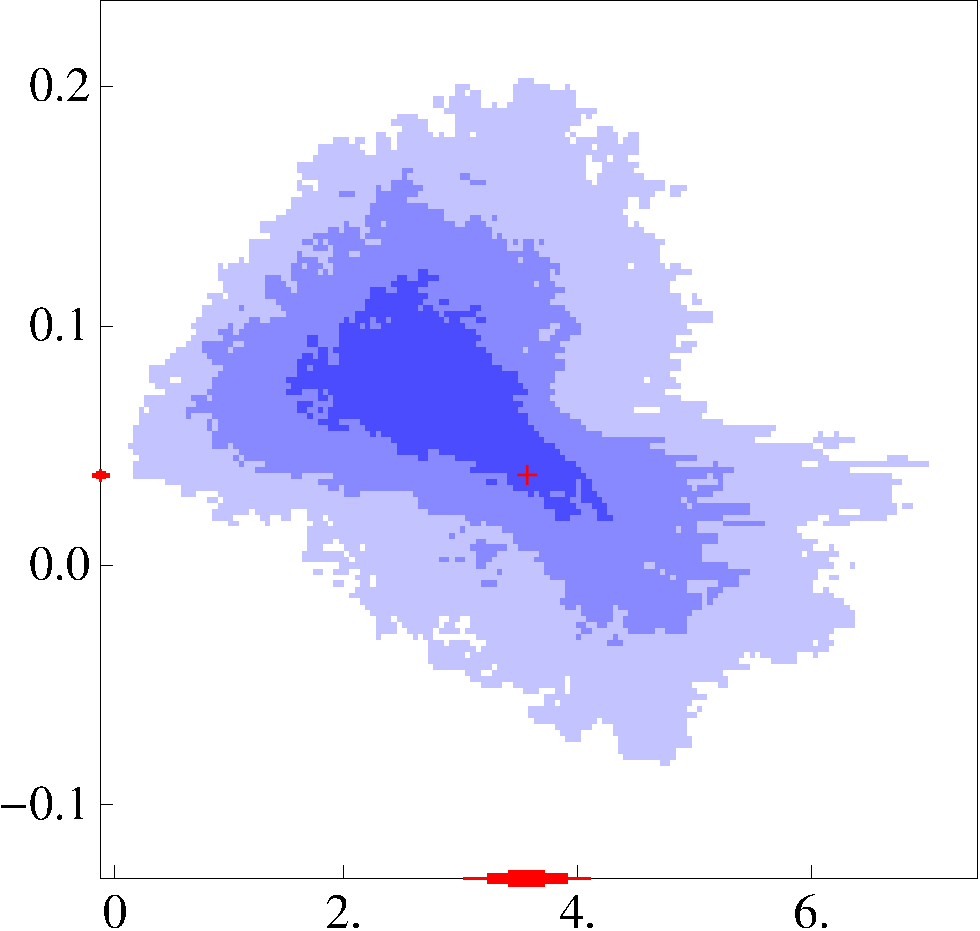}
\caption{$\AJPP$ vs. $\Bsmm\times 10^{9}$ in the \VL model (68\%, 95\% and 99\% CL regions, darker to lighter). Red bars (thicker to thinner) show the $1$, $2$ and $3$ $\sigma$ ranges corresponding to a SM fit. The experimental inputs are the bound $\Bsmm<4.5\times 10^{9}$ at 90\% CL and the value $\AJPP=0.002\pm 0.0873$.}\label{fig:AJPsiPhi:vs:Bsmumu:VL}
\end{center}
\end{figure}

\begin{figure}[H]
\begin{center}
\subfigure[$\asld\times 10^3$ vs. $\aslb\times 10^3$.\label{fig:asld:vs:aslb}]{\includegraphics[height=0.28\textheight]{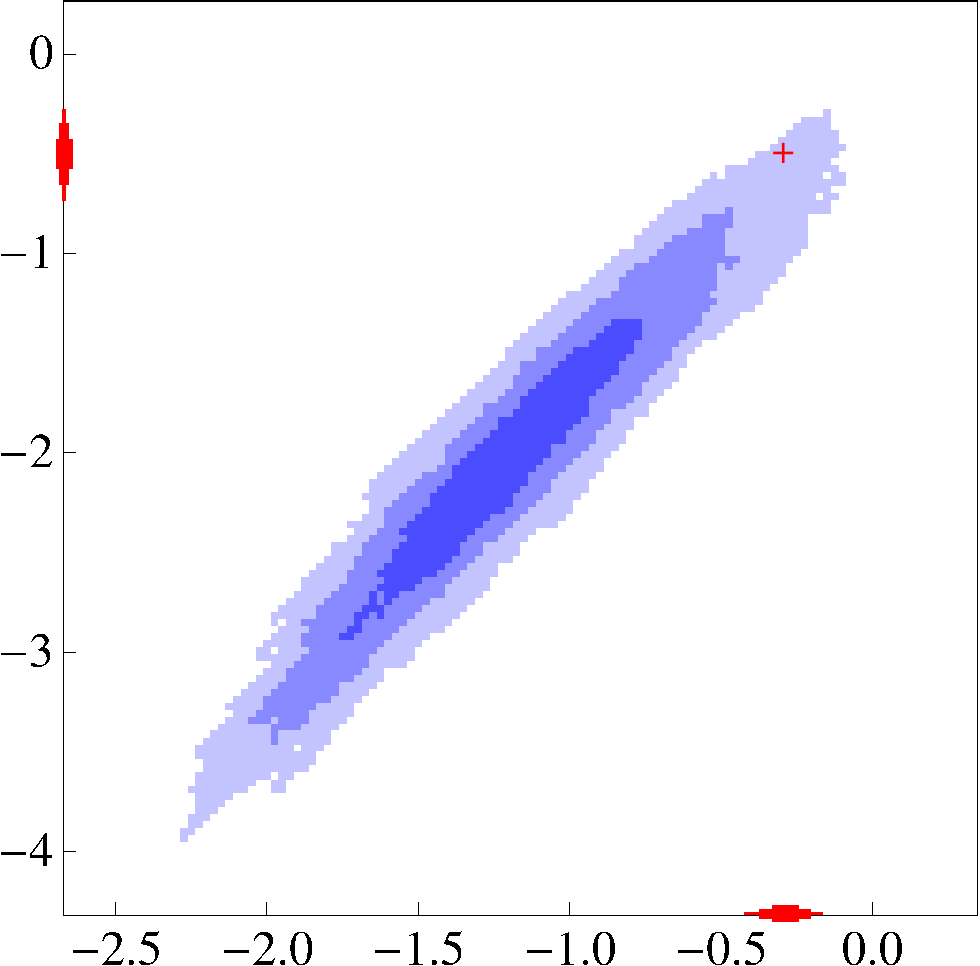}}\qquad
\subfigure[$\asld\times 10^3$ vs. $(\asldiff)\times 10^3$.\label{fig:fig:asld:vs:asldiff}]{\includegraphics[height=0.28\textheight]{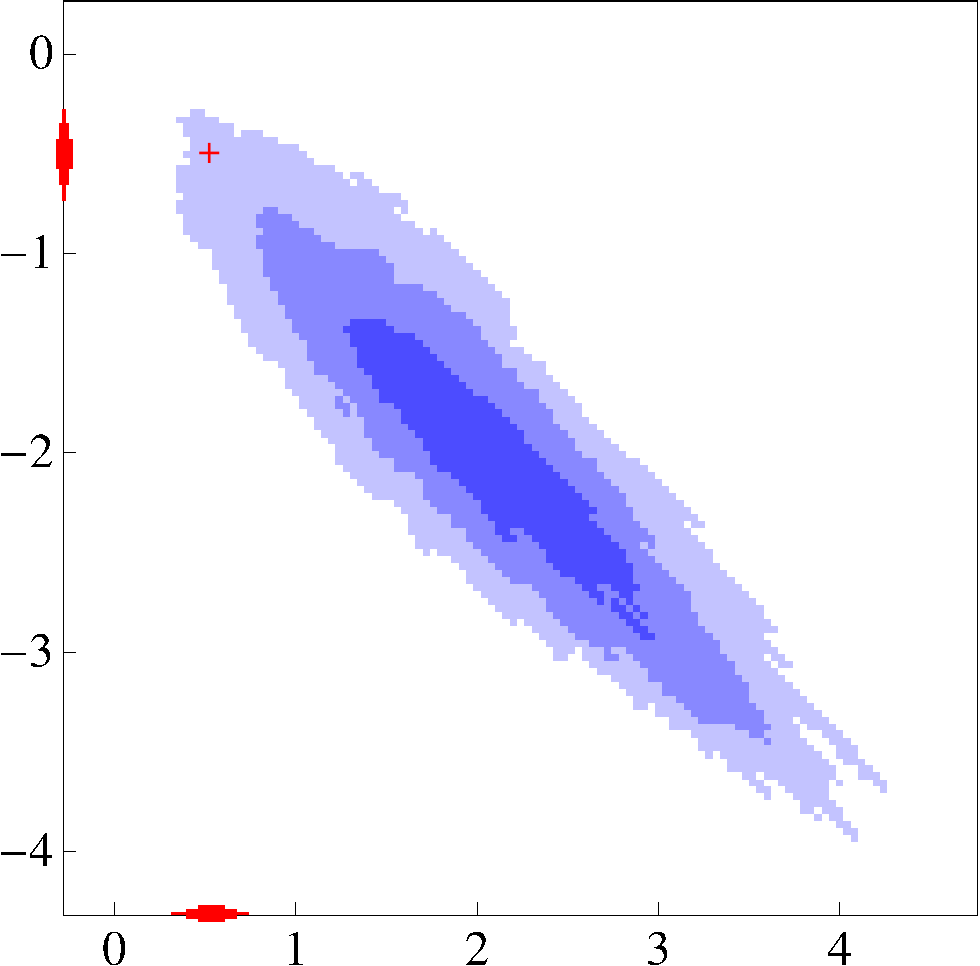}}
\caption{Correlations with $\asld$ in the \VL model, 68\%, 95\% and 99\% CL regions (darker to lighter). Red bars (thicker to thinner) show the $1$, $2$ and $3$ $\sigma$ ranges corresponding to a SM fit. The experimental value of $\asld$ is $\asld=(-3.0\pm 7.8)\times 10^{-3}$, $(\asldiff)$ is yet unmeasured.}\label{fig::asld:vs:var}
\end{center}
\end{figure}

In spite of the fact that there are not very large weak phases in the \BBsmix\ mixing, from the previous figure we must stress that values of $\asls$ and $\AJPP$ much larger than the SM value are allowed. These values are completely correlated. Another important correlation pointed out in \cite{Botella:2008qm} is among $\AJPP$ and $\beta_{s}$.
%%%%%
Note that in the SM this correlation --- for the chosen CP final eigenstate --- is positive, but in the \VL model it is negative. This was proven in reference \cite{Botella:2008qm} and therefore this correlation provides a consistency check of the entire analysis. It is shown in figure \ref{fig:betaS:vs:AJPsiPhi:VL}. In figure \ref{fig:AJPsiPhi:vs:Bsmumu:VL} we see that after the impressive bounds on the branching ratio $\Bsmm$ presented by LHCb \cite{Aaij:2012ac}, CMS \cite{Chatrchyan:2012rg} and ATLAS \cite{Aad:2012pn}, there is still room for values of $\AJPP$ of order 0.1\,.
%%%%%%%
%%%%%%%
It is apparent that smaller values --- below the SM value --- of $\Bsmm$ are more easily related to important deviations of $\AJPP$ from its SM value, especially towards 0.1\,.
In fig. \ref{fig::asld:vs:var} we can see two important results. First, the model cannot reproduce the value of $\aslb$ reported by the D0 collaboration, in spite of the fact that the model violates $3\times 3$ unitarity, a new ingredient generally not included in the majority of the model independent analysis. The maximum value allowed in this model for $\aslb$ is around $-2\times 10^{-3}$, a factor of \emph{seven} larger than the SM value but still at $3\sigma$ from the D0 reported value. 
Second, we also learn that the leading contribution in these models to the dimuon charge asymmetry comes from $\asld$, as it should after the
Lepton-Photon drama which implies a very strong indirect constraint on $\asls$.
For completeness we also give, in fig. \ref{fig:fig:asld:vs:asldiff}, the correlation among $\asld$ and the quantity which is expected that LHCb could deliver soon: $\asls-\asld$.
%%%%%%
%%%%%

\begin{figure}[H]
\begin{center}
\includegraphics[height=0.28\textheight]{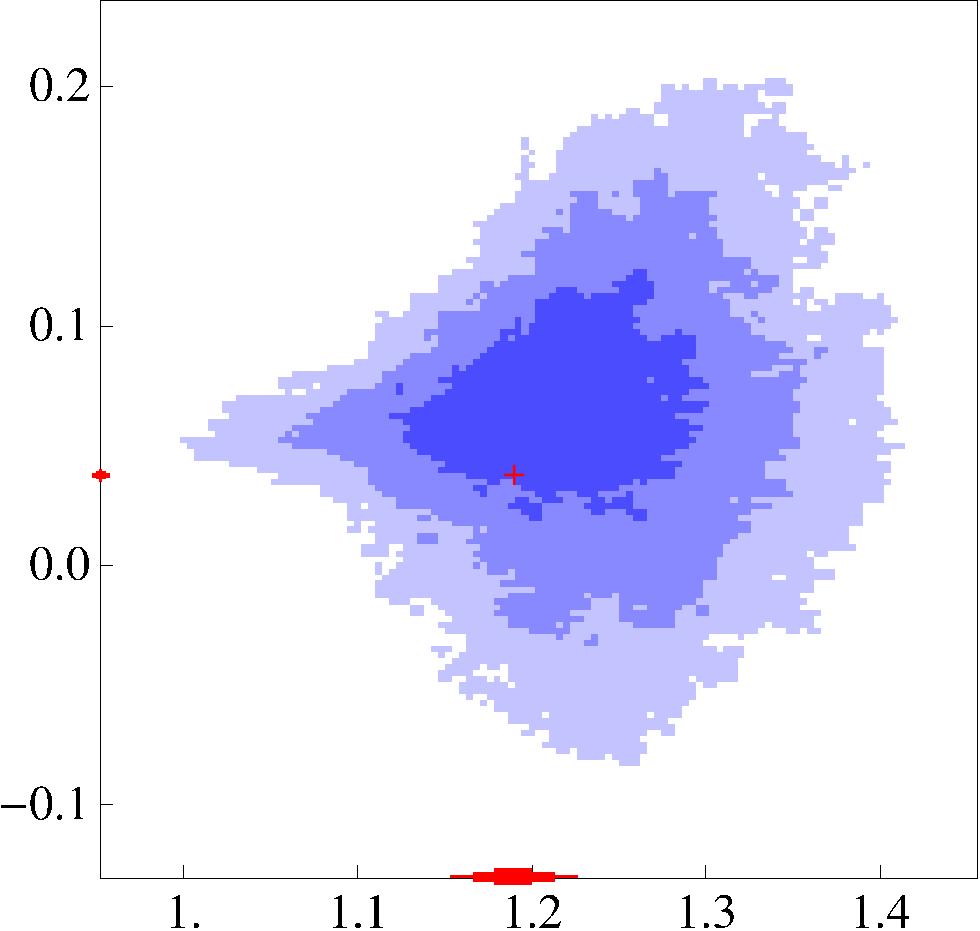}
\caption{$\AJPP$ vs. $\gamma$ in the \VL model, 68\%, 95\% and 99\% CL regions (darker to lighter). Red bars (thicker to thinner) show the $1$, $2$ and $3$ $\sigma$ ranges corresponding to a SM fit. The experimental values are $\AJPP=0.002\pm 0.0873$ and $\gamma=1.34\pm 0.24$.}\label{fig:AJPP:vs:gamma}
\end{center}
\end{figure}

The dominance of $\asld$ versus $\asls$ is again clear. 
In order to clarify the impact of future improvements on the measurement of $\gamma$, we show the correlation between $\AJPP$ and $\gamma$ in figure \ref{fig:AJPP:vs:gamma}.
It is clear that the largest departures of $\AJPP$ from the SM value can be easily accommodated in the region with larger $\gamma$.

\begin{figure}[h]
\begin{center}
\subfigure[$\Bsmm\times 10^9$ vs. $\BsG\times 10^4$.\label{fig:Bsmumu:vs:BsG}]{\includegraphics[height=0.28\textheight]{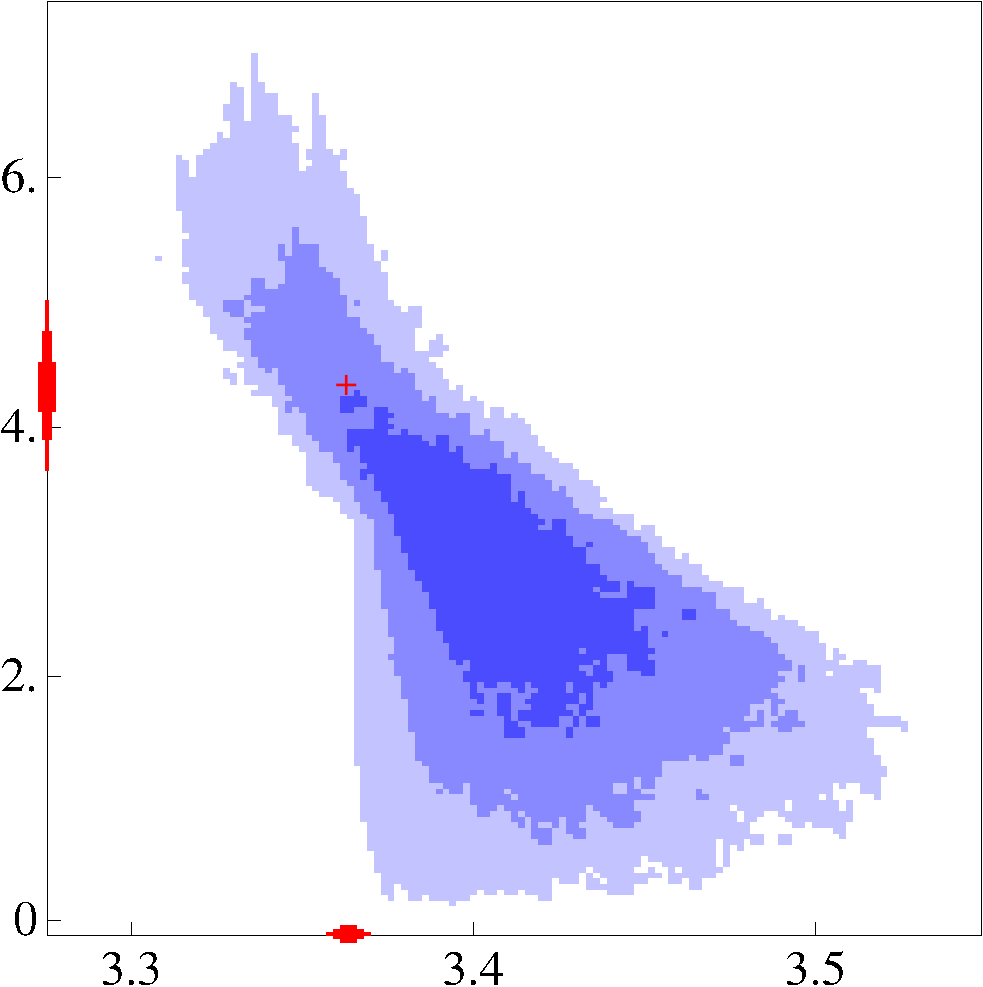}}\qquad
\subfigure[$\Bsmm\times 10^9$ vs. $\BXsmm\times 10^6$.\label{fig:Bsmumu:vs:BXsmumu}]{\includegraphics[height=0.28\textheight]{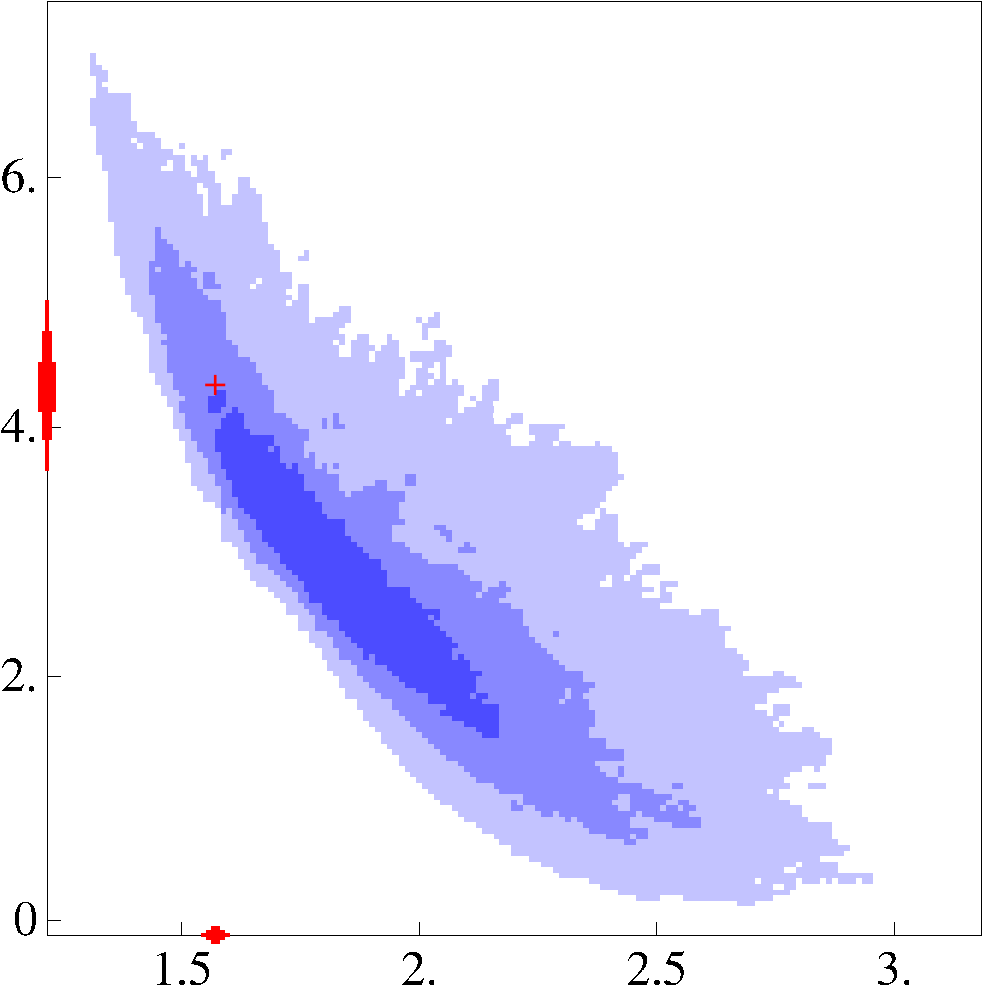}}
\caption{Correlations with $\Bsmm$ in the \VL model, 68\%, 95\% and 99\% CL regions (darker to lighter). Red bars (thicker to thinner) show the $1$, $2$ and $3$ $\sigma$ ranges corresponding to a SM fit. The experimental inputs are the bound $\Bsmm<4.5\times 10^{9}$ at 90\% CL and the values $\BsG=(3.56\pm 0.25)\times 10^{-4}$ and $\BXsmm=(1.60\pm 0.51)\times 10^{-6}$.}\label{fig:Bsmumu:vs:var}
\end{center}
\end{figure}

Interesting correlations in this model exist among $\Bsmm$ and $\BsG$ or $\BXsmm$. It is important to stress that low values of $\Bsmm$ are anticorrelated with large values of $\BsG$ and $\BXsmm$ as can be seen in fig. \ref{fig:Bsmumu:vs:var}.
%\clearpage
%%%%%%%
%%%%%%%
%%%%%%
%%%%%

\clearpage
% % % % % % % % % % %
%\section{The $\boldsymbol{ds}$ sector\label{sec:ds}} %{The $\boldsymbol{K^{0}}$ sector of sUVL}
\section{The ${ds}$ sector\label{sec:ds}} %{The $\boldsymbol{K^{0}}$ sector of sUVL}
% % % % % % % % % % %

Of course we have included the most restrictive kaon constraints in the analysis, both CP violating and CP conserving \cite{Cirigliano:2011ny}. As we have seen in previous sections, to eliminate the tensions in the $bd$ sector we need non negligible $\V{Td}$ and $\V{Tb}$. It seems that we can have significant NP contributions to $\KPinunu$ provided we have important $|\V{Td}\V{Ts}|$. It is worthwhile to look to the correlation among $\KPinunu$ and $|\V{Ts}|$, but also with $\AJPP$, that can be sizeable provided one has a significant $bs$ quadrangle.

%%%%%%
\begin{figure}[h]
\begin{center}
\subfigure[$\KPinunu\times 10^{10}$ vs. $|\Vfig{Ts}|$.\label{fig:KPinunu:vs:VTs}]{\includegraphics[height=0.28\textheight]{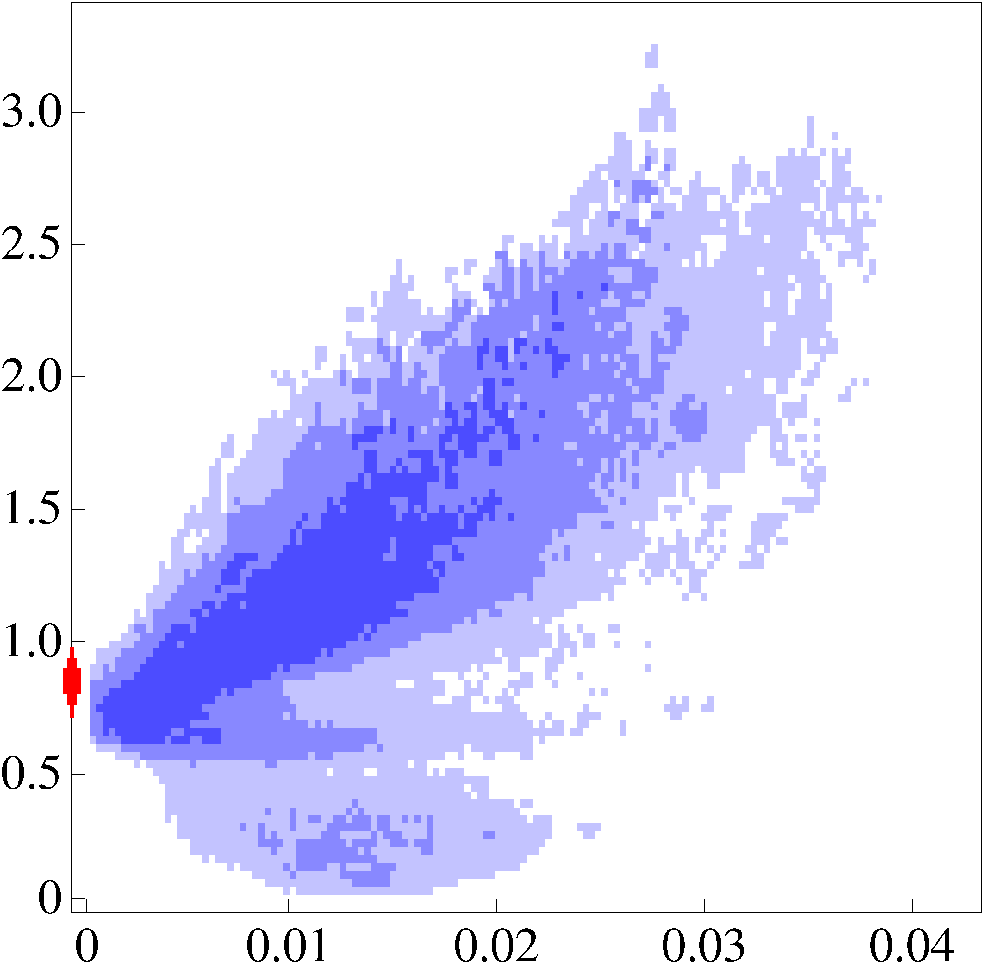}}\qquad
\subfigure[$\KPinunu\times 10^{10}$ vs. $\AJPP$.\label{fig:KPinunu:vs:AJPsiPhi}]{\includegraphics[height=0.28\textheight]{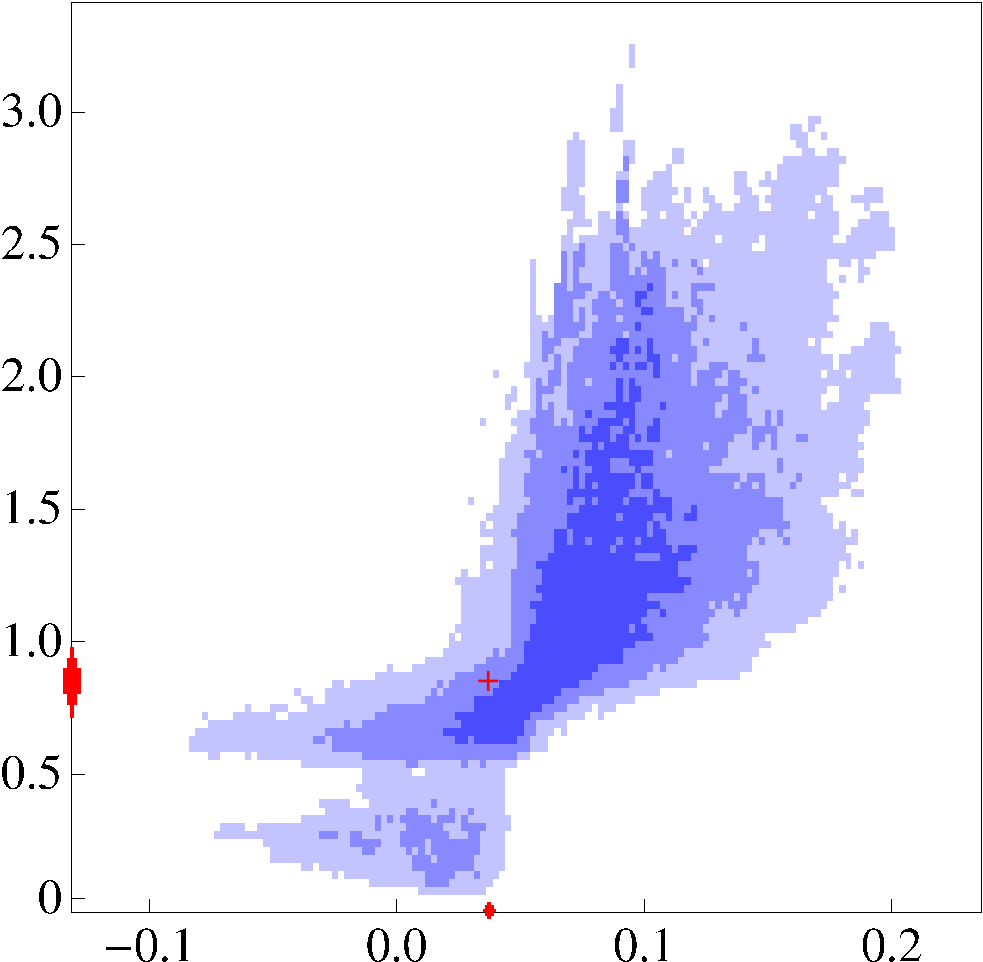}}
\caption{Correlations with $\KPinunu\times 10^{10}$ in the \VL model, 68\%, 95\% and 99\% CL regions (darker to lighter). Red bars (thicker to thinner) show the $1$, $2$ and $3$ $\sigma$ ranges corresponding to a SM fit. The experimental inputs are $\KPinunu=(1.73\pm 1.05)\times 10^{-10}$ and $\AJPP=0.002\pm 0.0873$.}\label{fig:KPinunu:vs:var}
\end{center}
\end{figure}
%%%%%

\begin{figure}[h]
\begin{center}
\includegraphics[height=0.28\textheight]{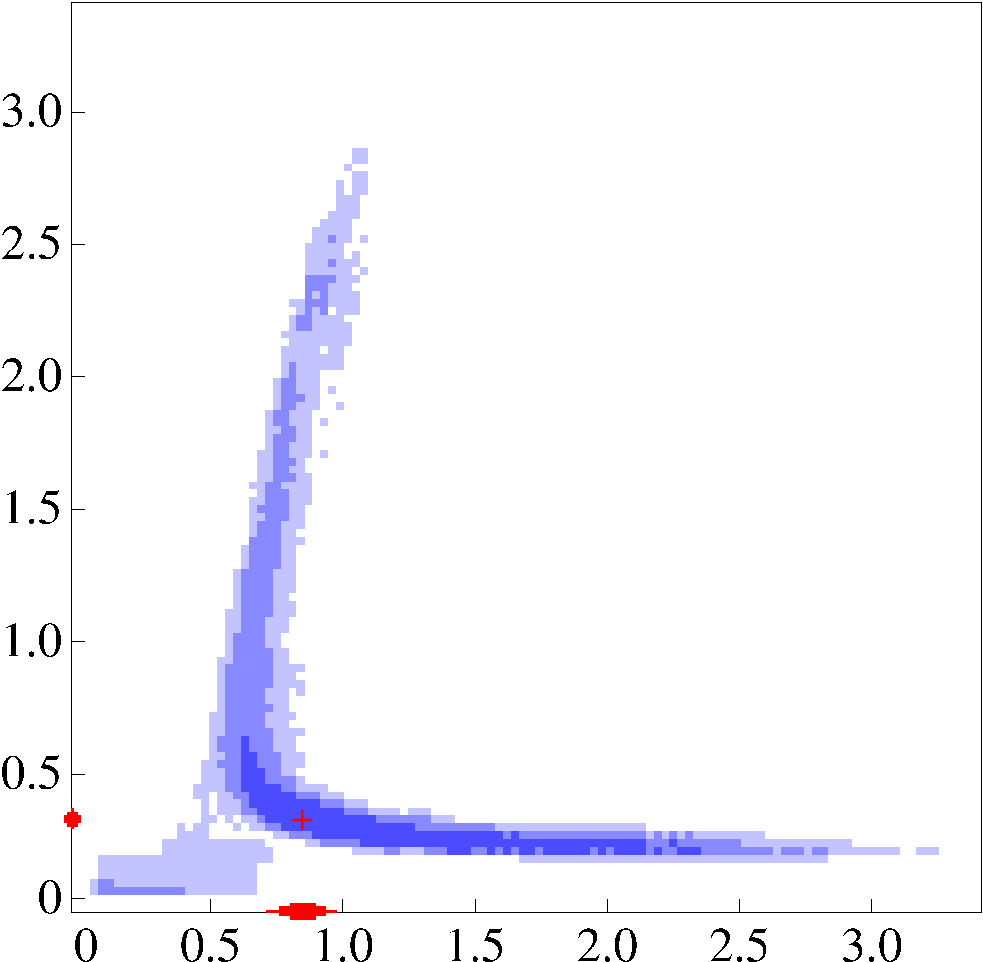}
\caption{$\KLPinunu\times 10^{10}$ vs. $\KPinunu\times 10^{10}$ in the \VL model, 68\%, 95\% and 99\% CL regions (darker to lighter). Red bars (thicker to thinner) show the $1$, $2$ and $3$ $\sigma$ ranges corresponding to a SM fit. The experimental inputs are $\KPinunu=(1.73\pm 1.05)\times 10^{-10}$ and the bound $\KLPinunu<2.8\times 10^{-8}$ at 90\% CL.}\label{fig:KLPinunu:vs:KPinunu}
\end{center}
\end{figure}

Both figures \ref{fig:KPinunu:vs:VTs} and \ref{fig:KPinunu:vs:AJPsiPhi} confirm our expectations. $\KPinunu$ clearly has a branch that grows with $|\V{Ts}|$ and we can see that positive values of $\AJPP$ --- larger than the SM value --- can accommodate values of the $\KPinunu$ enhanced with respect to the SM values by factors of two or more. It is important to point out that in this model the Grossman-Nir bound \cite{Grossman:1997sk} in the plane $\KPinunu$ vs. $\KLPinunu$ cannot be saturated even if the process $K_{L}\to \pi^{0}\nu\bar{\nu}$ can be enhanced by factors up to seven as it is shown in figure \ref{fig:KLPinunu:vs:KPinunu}.

%%%%%%%
%%%%%%%

%\clearpage
% % % % % % % % % % % % %
\section{The up sector\label{sec:up}}
% % % % % % % % % % % % %

% % % % % % % % % % % % %
\subsection{Rare top decays}
% % % % % % % % % % % % %

\begin{figure}[h]
\begin{center}
\includegraphics[height=0.28\textheight]{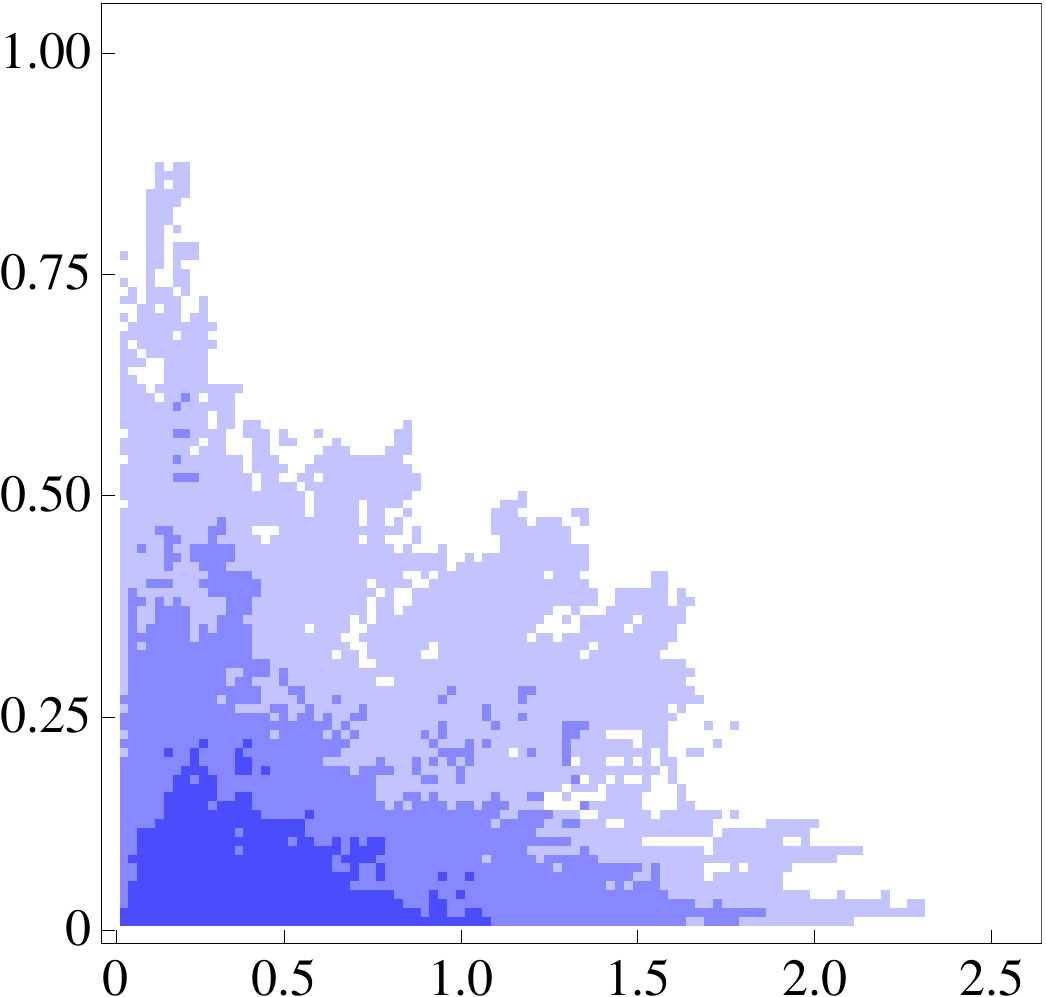}
\caption{$\tcZ\times 10^{5}$ vs. $\tuZ\times 10^{5}$ in the \VL model, 68\%, 95\% and 99\% CL regions (darker to lighter).}\label{fig:tcZ:vs:tuZ}
\end{center}
\end{figure}

A very distinctive signal of this class of models are the rare top decays. In this case, due to the presence of flavour changing neutral currents at tree level in the up sector, we have mainly $t\to cZ$ and $t\to uZ$ controlled by $|U_{c4} U_{t4}|$ and $|U_{u4} U_{t4}|$ respectively. Its natural order of magnitude is at the $10^{-6}$ level arriving in some cases to $10^{-5}$ as can be seen in the correlation plot in figure \ref{fig:tcZ:vs:tuZ}.

%%%%%%%
%%%%%%%

\begin{figure}[h]
\begin{center}
\subfigure[$\tcZ\times 10^{5}$.\label{fig:tcZ}]{\includegraphics[height=0.2\textheight]{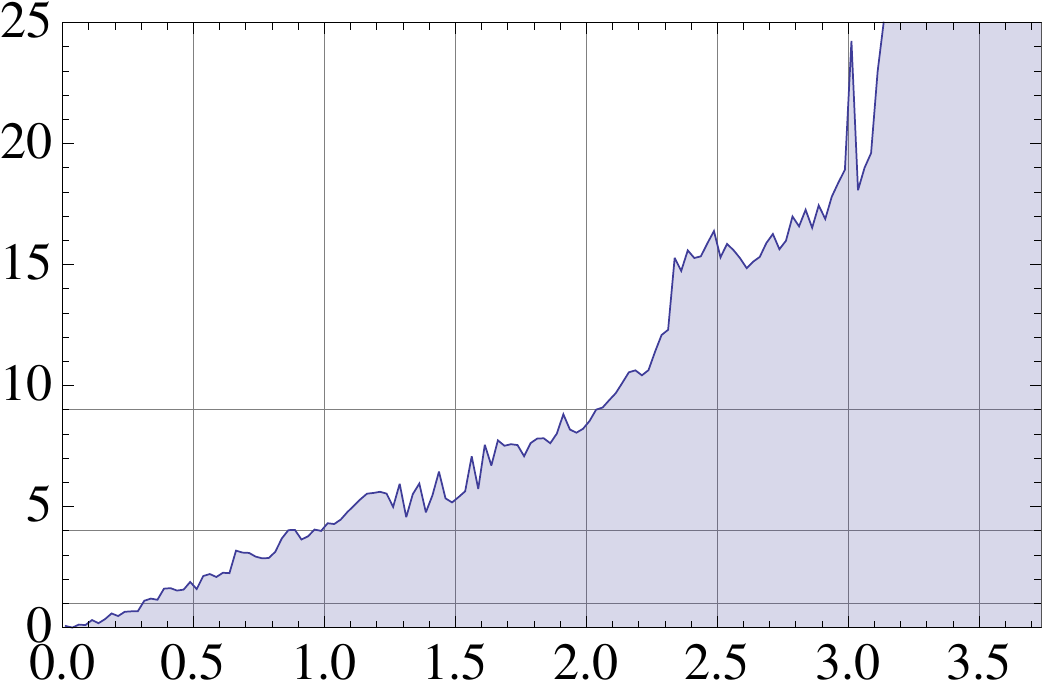}}\qquad
\subfigure[$\tuZ\times 10^{5}$.\label{fig:tuZ}]{\includegraphics[height=0.2\textheight]{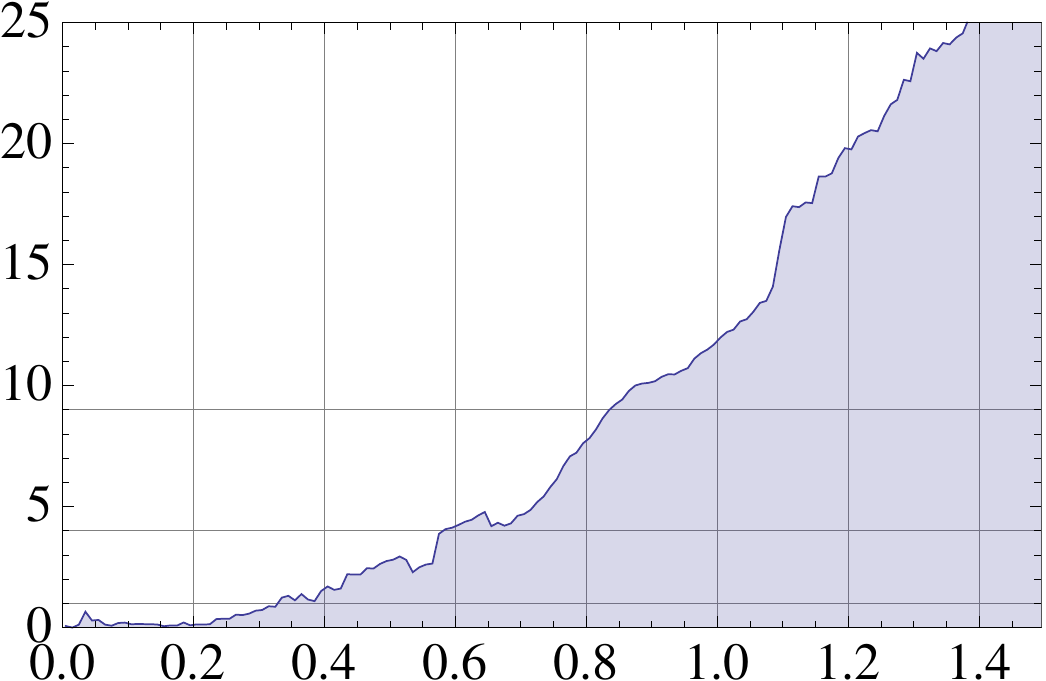}}
\caption{$\tcZ$ and $\tuZ$, $\Delta\chi^2$ profiles in the \VL model.}\label{fig:tqZ}
\end{center}
\end{figure}

We also display the individual likelihood profiles to see clearly their natural order of magnitude. With the values shown in figure \ref{fig:tqZ}, although very large compared to the SM values, it is going to be difficult to discover this class of NP using these rare top decays.

%%%%%
%%%%%

%\clearpage
% % % % % % % % % % % % %
\subsection{Charm}
% % % % % % % % % % % % %

\begin{figure}[h]
\begin{center}
\subfigure[$\Dmm\times 10^{10}$\label{fig:Dmumu}]{\includegraphics[height=0.2\textheight]{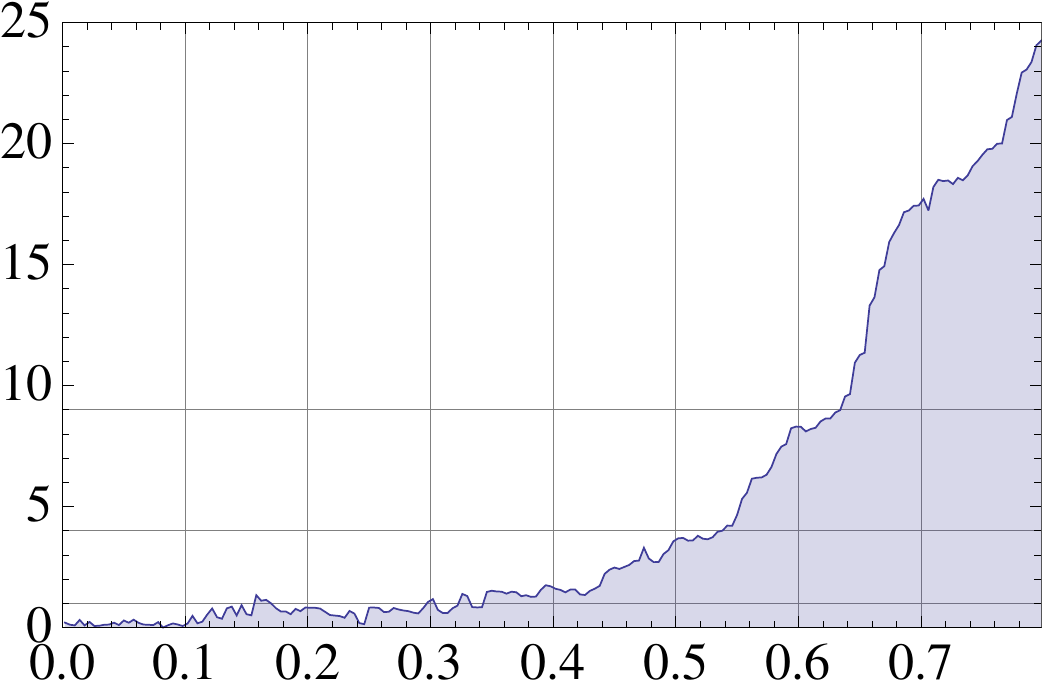}}\qquad
\subfigure[$\xD\times 10^{2}$\label{fig:xD}]{\includegraphics[height=0.2\textheight]{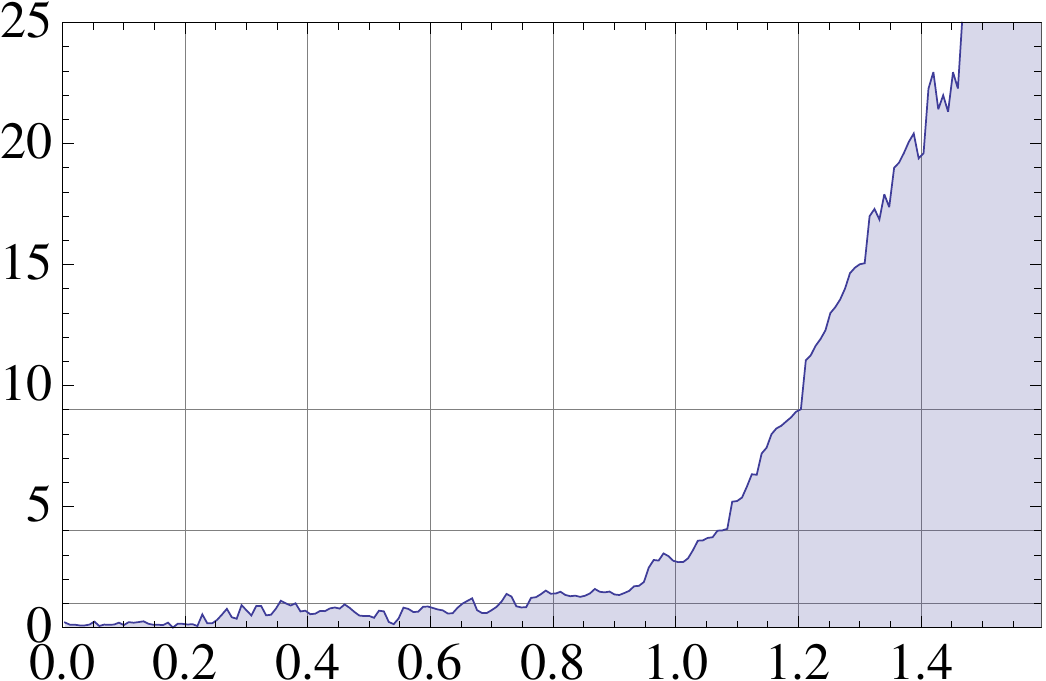}}
\caption{$\Dmm$ and $\xD$, $\Delta\chi^2$ profiles in the \VL model.}\label{fig:charm}
\end{center}
\end{figure}

With the presence of FCNC at tree level in the up sector, the \DDmix\ mixing can be explained naturally in the framework of this model. The mixing parameter $x_{D}$ has been used to bound the intensity of these FCNC, still SM long distance contributions can be dominant \cite{Golowich:2006gq}.
Automatically we get the short distance leading contribution to $\Dmm$ \cite{Golowich:2009ii}. The results are presented in figure \ref{fig:charm}; as we can see, the natural range of the branching ratio for $\Dmm$ is of the order of $10^{-11}$.
Recently other surprising results in the charm sector have arised, as is the case of the direct CP violation in the channels $D^{0}\to \pi^{+}\pi^{-}$ and/or $D^{0}\to K^{+}K^{-}$ \cite{Aubert:2007if,Staric:2008rx,Aaltonen:2011se,Aaij:2011in}. The new tree level FCNC contribution has a suppression similar to the one we observe in $\Dmm$, so we do not expect that \VL FCNC can give an important contribution to this direct CP violation. Other mechanisms where vector-like quarks can play a role are at present under scrutiny \cite{Delaunay:2012cz}.

%%%%%
%%%%%

%\clearpage
% % % % % % % % % % % %
%\subsection{Additional correlations and $\boldsymbol{\mT}$} %{Other correlations and the value of $\mT$}
\subsection{Additional correlations and ${\mT}$} %{Other correlations and the value of $\mT$}
\begin{figure}[h]
\begin{center}
\subfigure[$\mT$ (GeV) vs. $|\Vfig{Td}|$.\label{fig:VTd:vs:mT}]{\includegraphics[height=0.28\textheight]{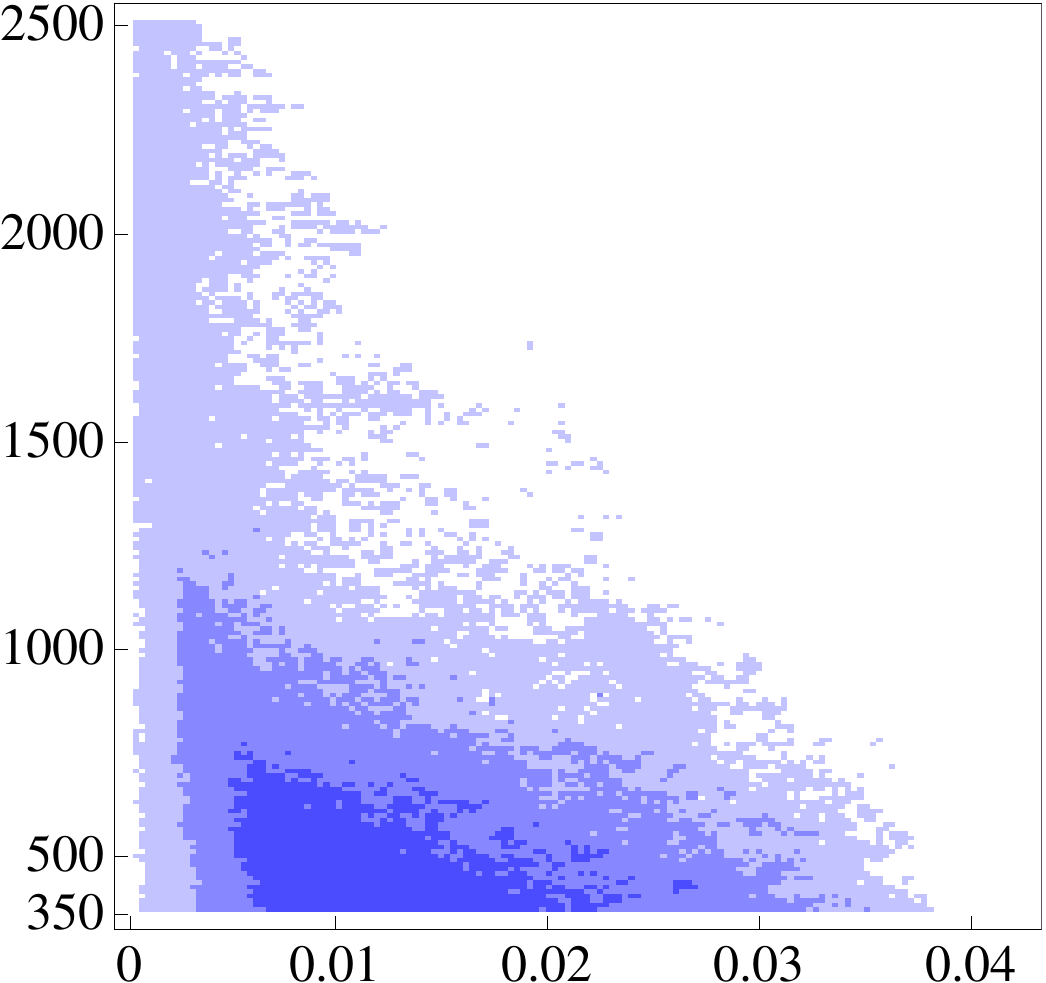}}\qquad
\subfigure[$\mT$ (GeV) vs. $|\Vfig{Ts}|$.\label{fig:VTs:vs:mT}]{\includegraphics[height=0.28\textheight]{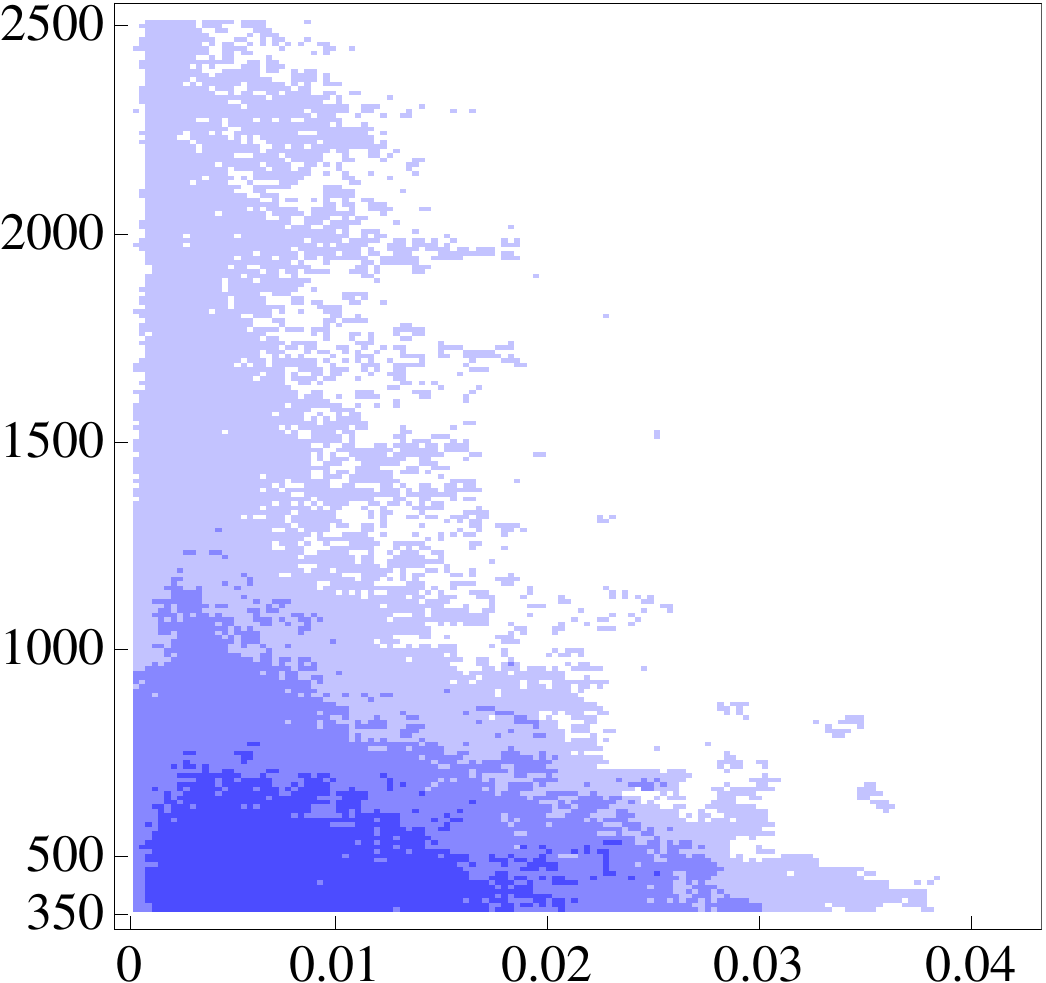}}\\
\subfigure[$\mT$ (GeV) vs. $|\Vfig{Tb}|$.\label{fig:VTb:vs:mT}]{\includegraphics[height=0.28\textheight]{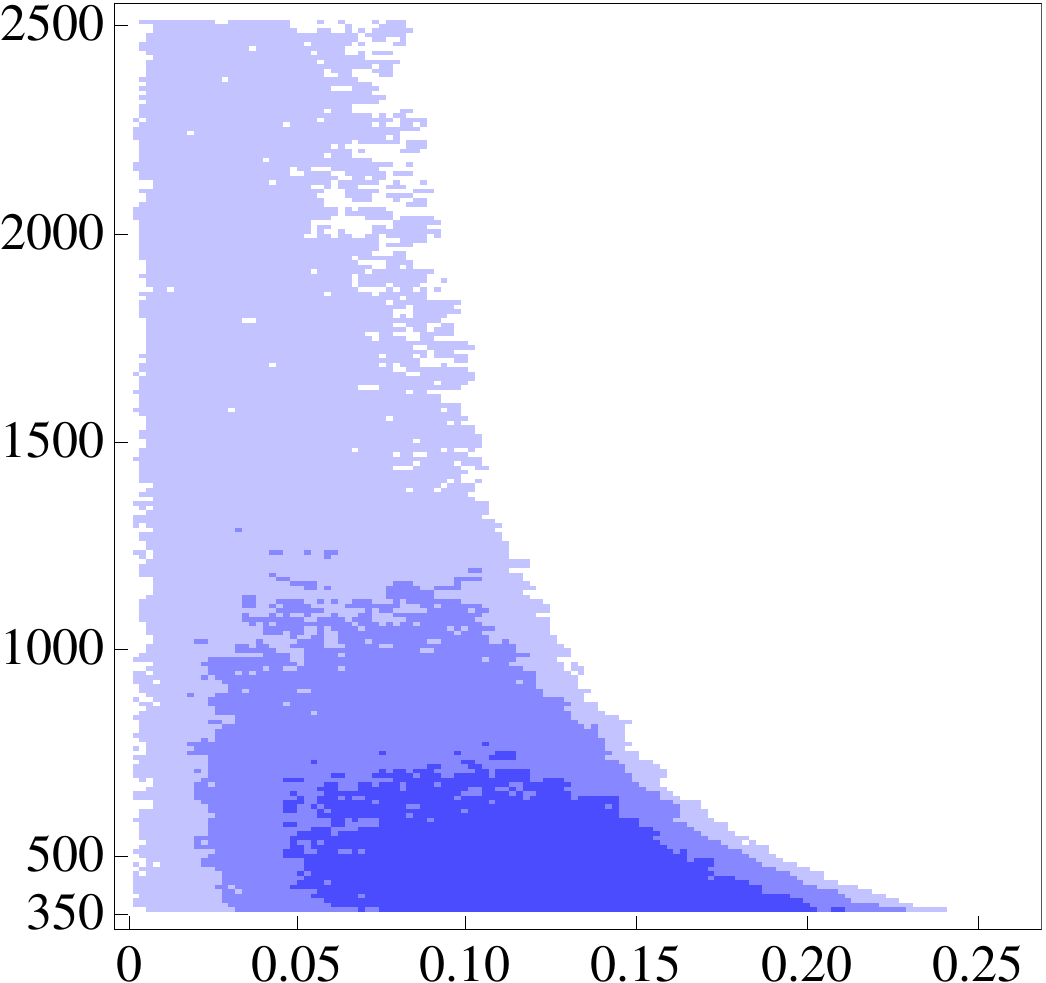}}
\caption{Correlations between $\mT$ and different $|\Vfig{Tq}|$ in the \VL model, 68\%, 95\% and 99\% CL regions (darker to lighter).}\label{fig:VTq:vs:mT}
\end{center}
\end{figure}

% % % % % % % % % % % %
A very important issue in these models is the presence of a new non-sequential up type quark. This new quark decouples from the low energy theory as soon as its mass grows. This decoupling, in particular, reflects the fact that in this limit all its couplings to light quarks vanish simultaneously. The consistency of the fit with this picture can be seen in the plots in figure \ref{fig:VTq:vs:mT}.

%%%%%
%%%%%

\begin{figure}[h]
\begin{center}
\includegraphics[height=0.28\textheight]{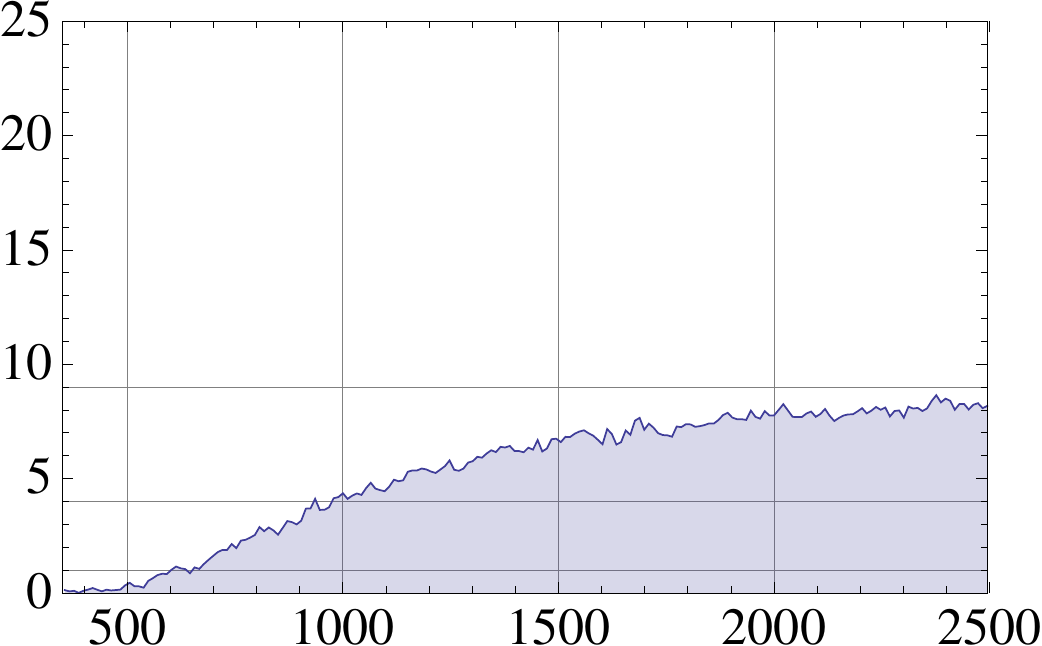}
\caption{$\mT$ (GeV) $\Delta\chi^2$ profile in the \VL model.}\label{fig:mT}
\end{center}
\end{figure}

We remind that recently D0 \cite{Abazov:2010ku}, CMS and ATLAS have published bounds on vector-like quark masses with values that range from 475 GeV in the CMS case \cite{Chatrchyan:2011ay} to 900 GeV in the ATLAS one \cite{Aad:2011yn}. It has to be stressed that in all these cases strong assumptions are made about the couplings of gauge bosons to this heavy quark and a light one. In particular CMS assumes that $T\to tZ$ is the dominant decay channel (100\% branching ratio). In our case it is more likely to have a branching ratio around $25\%$, which reduces the cross section by a substantial amount. In the ATLAS case it is implicitly assumed that $|\V{Td}| \sim |\V{Ts}|\geq 10^{-1}$, that is at least a factor of ten larger than what flavour data allows; the corresponding cross section has to be reduced accordingly.
It is out of the scope of this paper to analyse in detail the $\mT$ lower bounds from direct production, but following the previous comments, it is clear that $T$ masses as light as 350 GeV have to be considered.
The most important result of our analysis is the likelihood profile for $\mT$, shown in figure \ref{fig:mT}.

%%%%%
%%%%%

\begin{figure}[h]
\begin{center}
\includegraphics[height=0.28\textheight]{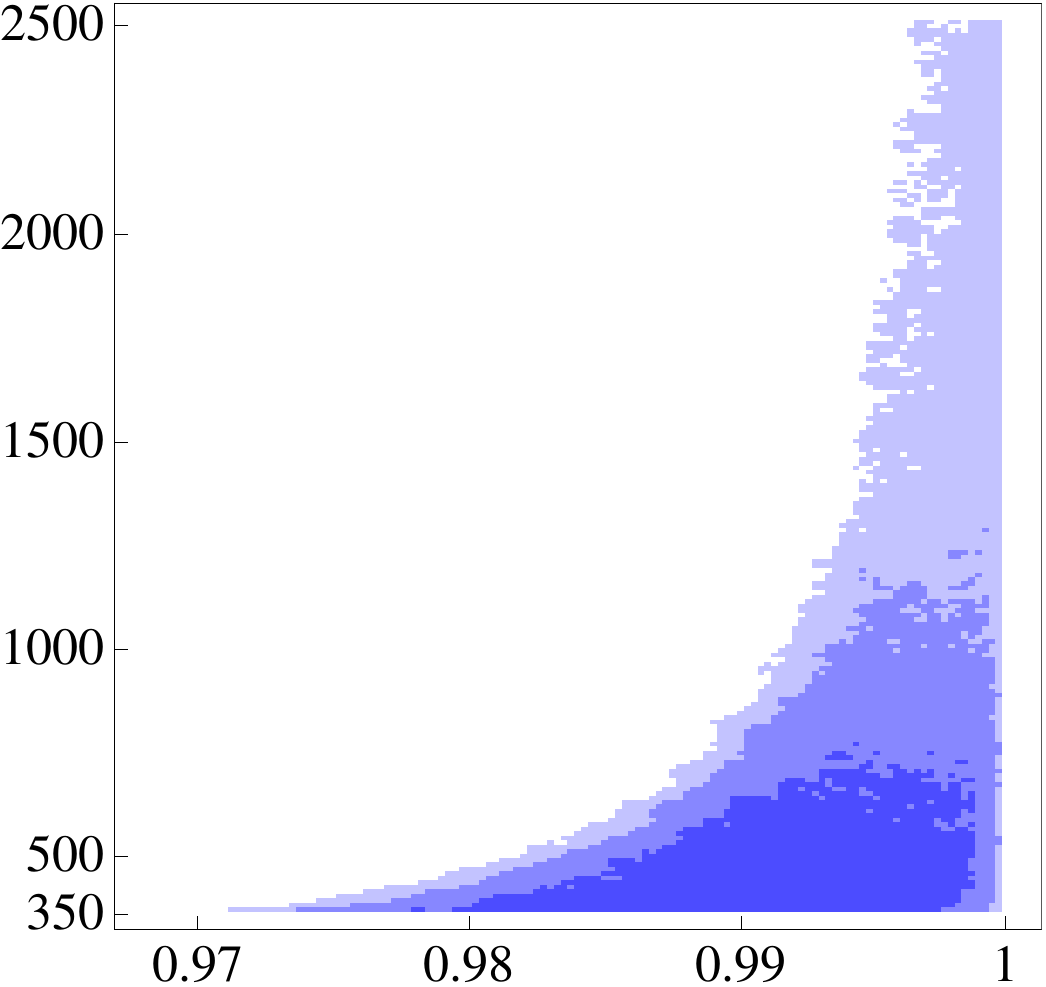}
\caption{$\mT$ (GeV) vs. $|\Vfig{tb}|$ in the \VL model, 68\%, 95\% and 99\% CL regions (darker to lighter).}\label{fig:mT:vs:Vtb}
\end{center}
\end{figure}

\begin{figure}[h]
\begin{center}
\includegraphics[height=0.28\textheight]{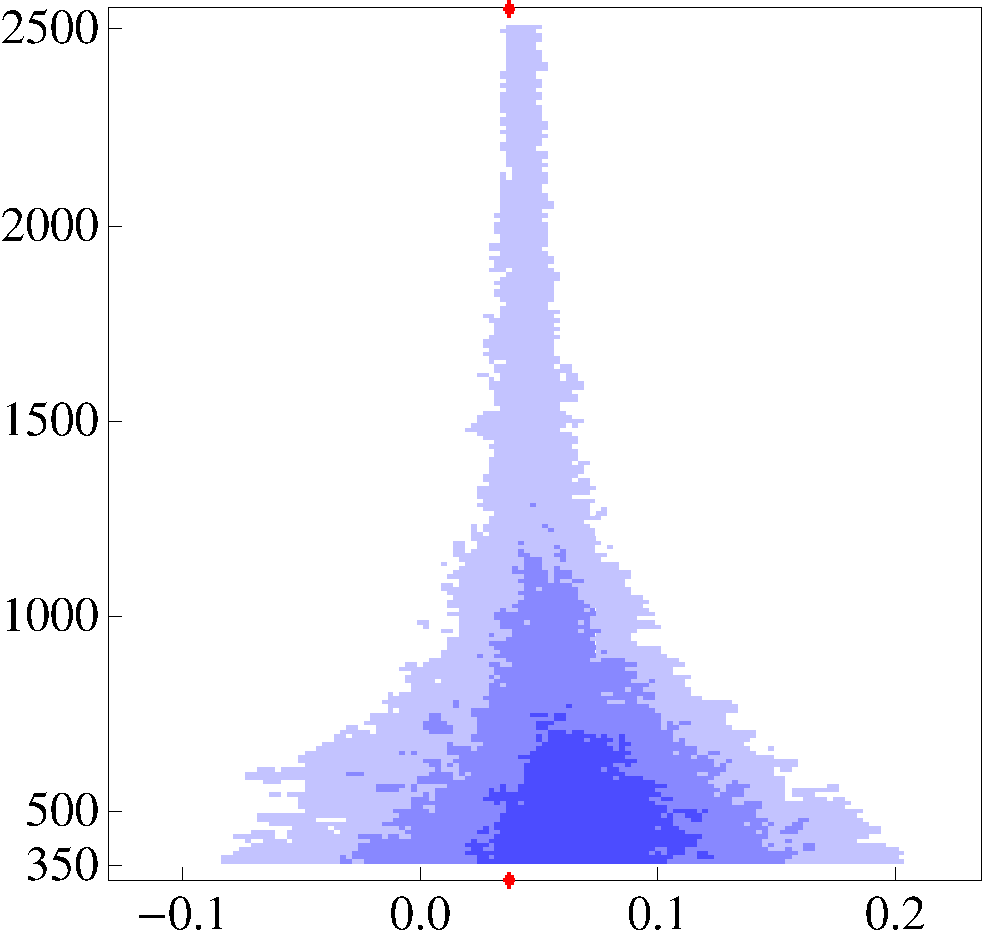}
\caption{$\mT$ (GeV) vs. $\AJPP$ in the \VL model, 68\%, 95\% and 99\% CL regions (darker to lighter). Red bars (thicker to thinner) show the $1$, $2$ and $3$ $\sigma$ ranges corresponding to a SM fit. The experimental value is $\AJPP=0.002\pm 0.0873$.}\label{fig:mT:vs:AJPsiPhi}
\end{center}
\end{figure}

\begin{figure}[h]
\begin{center}
\subfigure[$\mT$ (GeV) vs. $\asld$.\label{fig:mT:vs:ASLd}]{\includegraphics[height=0.28\textheight]{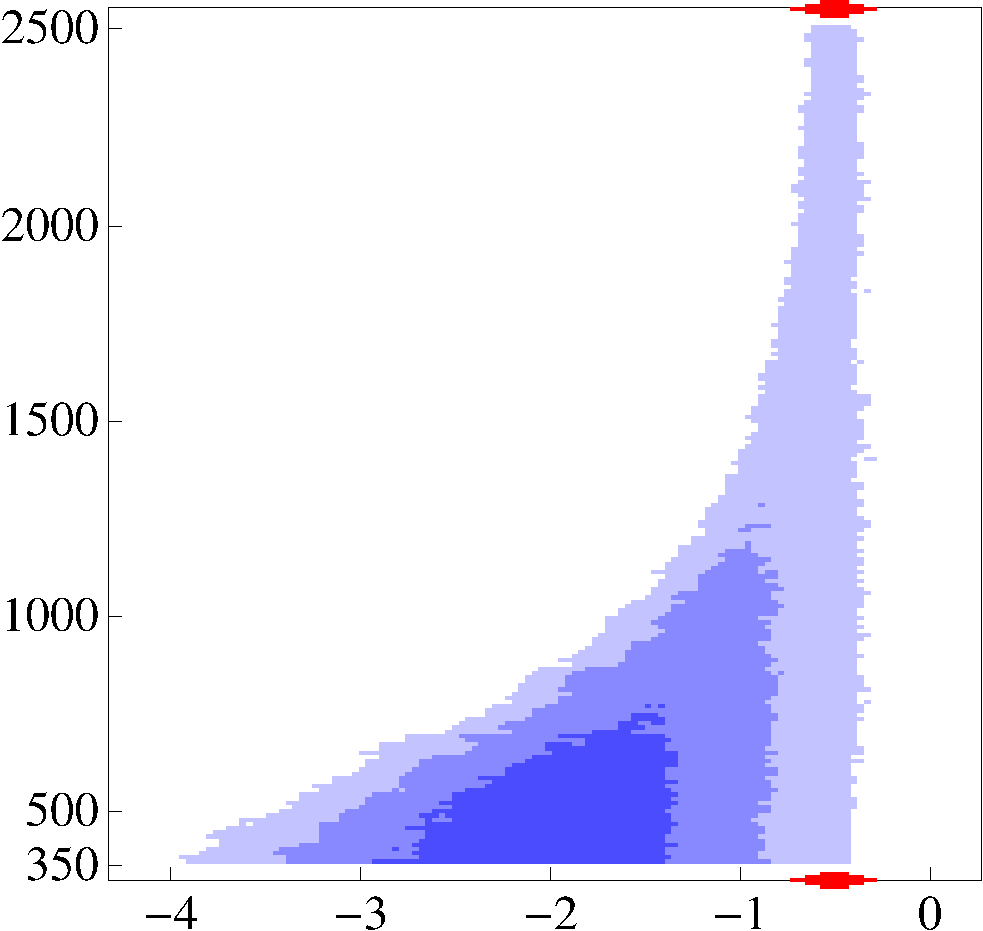}}\qquad
\subfigure[$\mT$ (GeV) vs. $\asldiff$.\label{fig:mT:vs:ASLdiff}]{\includegraphics[height=0.28\textheight]{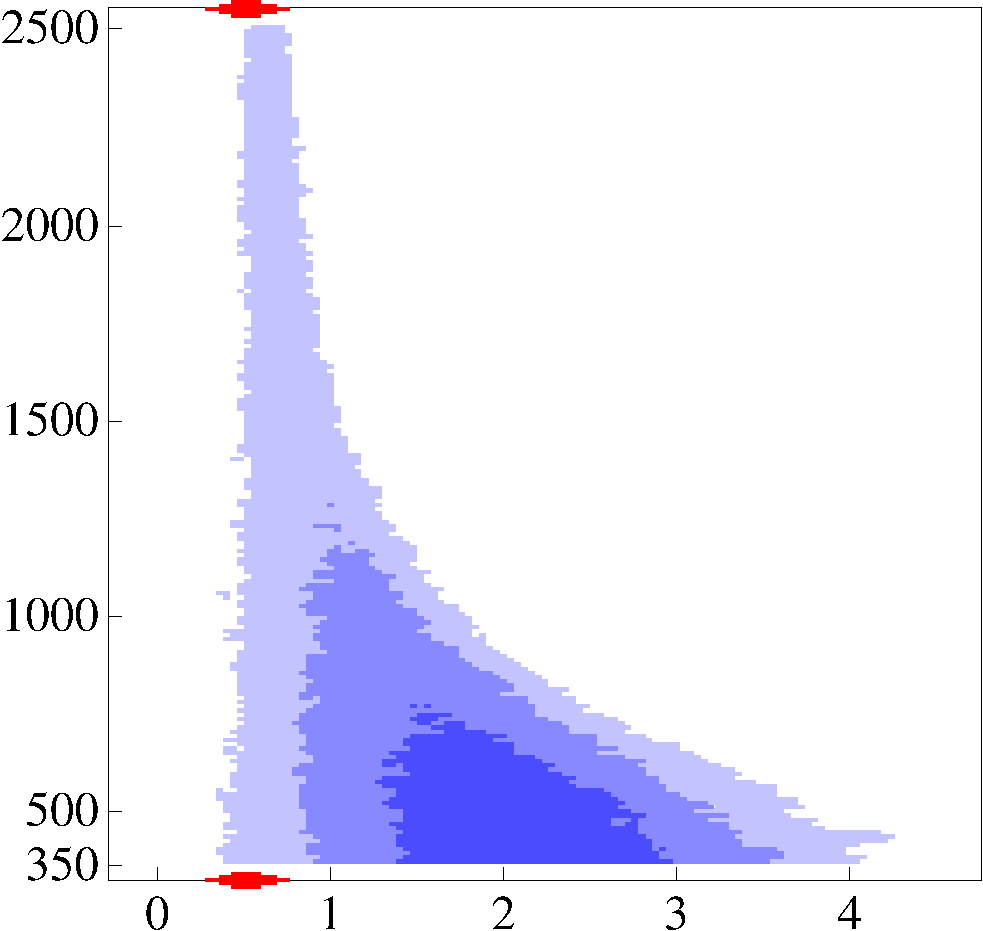}}
\caption{$\mT$ (GeV) vs. semileptonic asymmetries in the \VL model, 68\%, 95\% and 99\% CL regions (darker to lighter). Red bars (thicker to thinner) show the $1$, $2$ and $3$ $\sigma$ ranges corresponding to a SM fit. The experimental value is $\asld=(-3.0\pm 7.8)\times 10^{-3}$.}\label{fig:mT:vs:ASLvars}
\end{center}
\end{figure}

\begin{figure}[h]
\begin{center}
\subfigure[$\mT$ (GeV) vs. $\BsG$.\label{fig:mT:vs:BsG}]{\includegraphics[height=0.28\textheight]{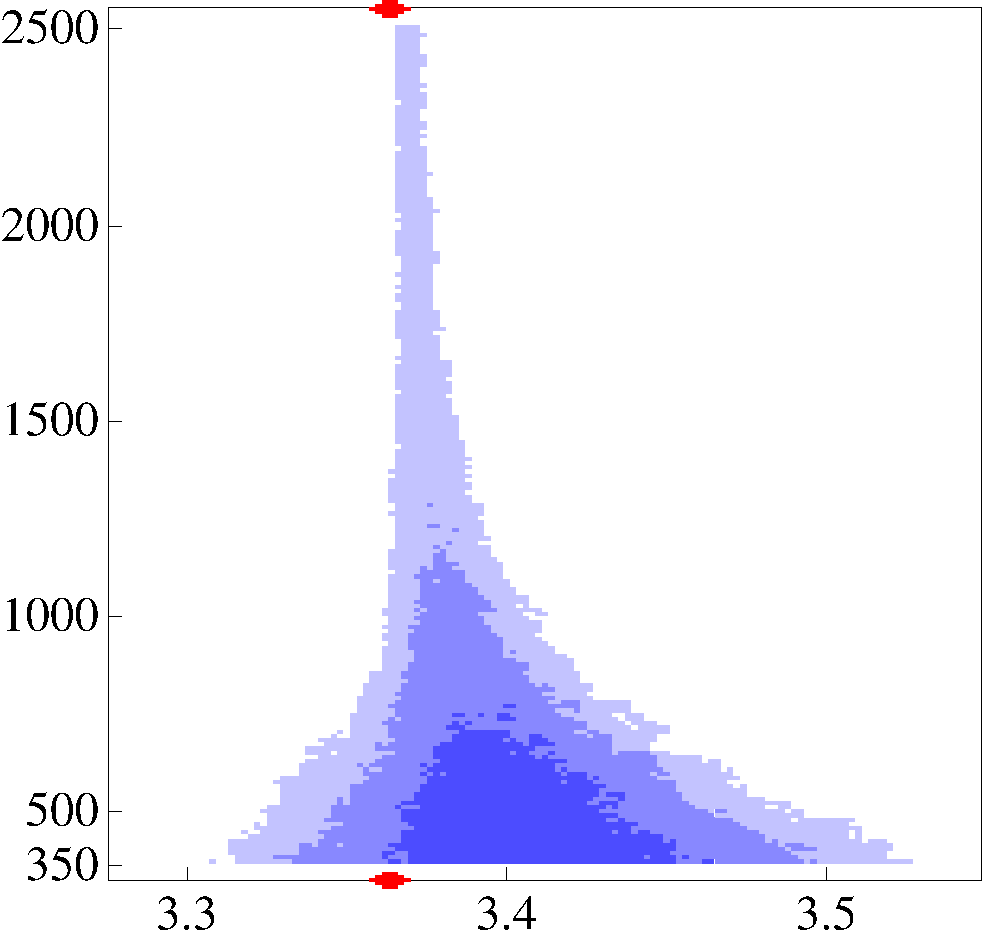}}\qquad
\subfigure[$\mT$ (GeV) vs. $\BTNu$.\label{fig:mT:vs:BTauNu}]{\includegraphics[height=0.28\textheight]{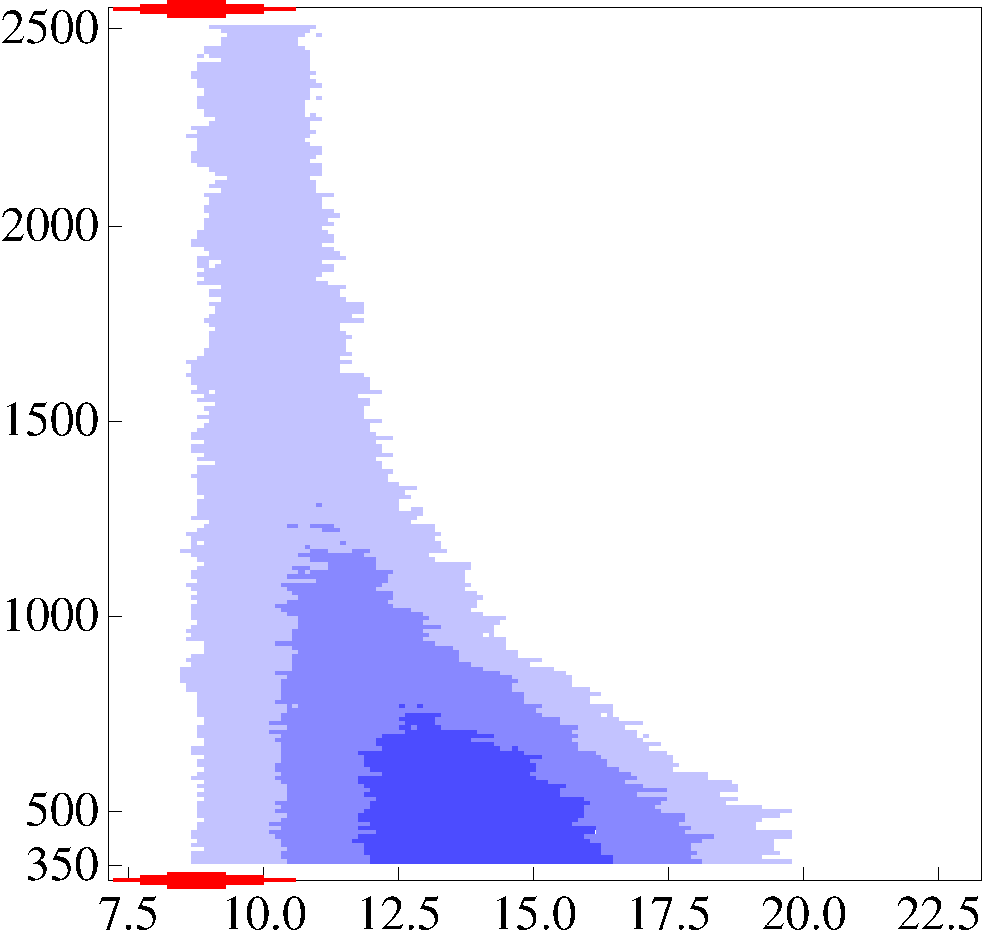}}
\caption{$\mT$ (GeV) correlations in the \VL model, 68\%, 95\% and 99\% CL regions (darker to lighter). Red bars (thicker to thinner) show the $1$, $2$ and $3$ $\sigma$ ranges corresponding to a SM fit. The experimental values are $\BsG=(3.56\pm 0.25)\times 10^{-4}$ and $\BTNu=(16.8\pm 3.1)\times 10^{-5}$.}\label{fig:mT:vs:BsG:BTNu}
\end{center}
\end{figure}

We can conclude that at 68\% CL, $m_{T}<0.6$ TeV and at 95\% CL $\mT<1$ TeV. The reason of having upper bounds for these mass values is that vector like quarks decouple, so in the infinite singlet mass limit this model should reproduce the SM and cannot solve the known tensions. With a heavy new top $T$ relatively light, in addition to an expected early production at the LHC, there will be other interesting signals: some of them will
be also measured at the LHC. Several examples can be seen in the following plots. The oblique parameter $\Delta T$ controls figure \ref{fig:mT:vs:Vtb}, that clearly says that an important departure of $\left\vert \V{tb}\right\vert $ from $1$ will point to the presence of a light
vector-like quark. The same happens with a sizeable departure from the SM
value of several observables that we present in plots \ref{fig:mT:vs:AJPsiPhi}, \ref{fig:mT:vs:ASLvars} and \ref{fig:mT:vs:BsG:BTNu}.

%%%%%
%%%%%

%%%%%
%%%%%

%%%%%
%%%%%

%%%%%
%%%%%

$\AJPP$, $\asld$ and $\asldiff$ can get values
out of the SM range more easily for a relatively light $T$ quark. $\BTNu$ and $\BsG$ can adjust better to its
experimental values with a light $m_{T}$.

\clearpage
\section{Conclusions\label{sec:conclusions}}

We have analysed most of the flavour data in the quark sector in the framework of one of the simplest extensions of the SM. We have incorporated an additional up quark of vector like character and singlet under $SU(2)_{L}$. Its existence leads to the presence of FCNC and deviations of $3\times 3$ unitarity, incorporated in a great variety of much more sophisticated models such as extra dimensions, little Higgs, etc \cite{Antoniadis:1994yi,Pomarol:1998sd,delAguila:2000aa,delAguila:2000rc,Agashe:2004ay,Perelstein:2005ka,Casagrande:2008hr,Blanke:2009am,Buras:2009ka,Bigi:2009df}. 
We have started by looking again to a model independent analysis to stress the fact that in the down sector the most important tensions appear in the $bd$ sector. We realize that in the \VL model, despite the violation of $3\times 3$ unitarity, the situation is similar. In fact it is with the change of a unitary $bd$ triangle to a quadrangle that we can improve the tensions among $\BTNu$ and $\AJPKs$. By the same token, an enlargement of $|\V{ub}|$ --- improving also $\AJPKs$ --- implies an enhancement of $\asld$ in the right direction of the $D0$ result of $\aslb$, but not enough. Also $\Bdmm$ can get  larger values compared to the SM when $\BTNu$ is in the right place.

In the $bs$ sector, in spite of the important constraints on mixing and CP violation coming from LHCb, it is still possible to have larger than SM values by some factors for  $\asls$ and $\AJPP$. Interestingly enough, larger values of $\AJPP$ are correlated with lower than SM values for $\Bsmm$ --- but not too small ---. In this respect, for this model, the rare decay $\Bdmm$ seems more critical than the companion $\Bsmm$. It is also interesting to remark that if a deviation of the SM value of $\AJPP$ is found, then $\gamma$ must be relatively large compared to its SM value. For the expected LHCb measurement, $\asldiff$, we get natural values around $\left(2-3\right) \times 10^{-3}$, a factor of five larger than the SM one. Needless to say, although we get a huge enhancement for $\aslb$ still we are at more than $2\sigma$ from the D0 result. In the kaon sector, still an important enhancement of a factor of seven is possible in $\KLPinunu$.
A very distintictive signature of this model is the enormous enhancement of the rare top decays $t\to Zq$ (FCNC). It is $t\to Zc$ that gets the largest values, of the order of $10^{-5}$. Similar to these processes is $\Dmm$, that can reach values of $5\times 10^{-11}$. Finally we have to stress that the fit prefers a heavy top below $600(1000)$ GeV at $68\% (95\%)$ CL. With this light up vector like quark we expect important deviations from the SM values of different observables.

The flavour sector of the SM is being tested at an unprecedented level of accuracy and even more stringent tests will be available in the near future. In this paper we have studied how some of the present tensions between the SM and experimental data can be alleviated in the framework of a simple extension of the SM, where a up-type vector like quark is added to the SM. Various correlations among physical observables are derived and provide a crucial set of experimental tests of the model.
%\clearpage

% % % % % % % % % % % % % % % % % % % % % % % % % % %
\acknowledgments
% % % % % % % % % % % % % % % % % % % % % % % % % % %
\indent This work was partially supported by \emph{Funda\c{c}\~ao para a Ci\^encia e a Tecnologia} (FCT, Portugal) through the projects CERN/FP/109305/2009, CERN/FP/116328/2010 and\newline CERN/FP/123580/2011, PTDC/FIS/098188/2008 and CFTPFCT Unit 777 which are partially funded through POCTI (FEDER), by Marie Curie Initial Training Network UNILHC PITN-GA-2009-237920, by \emph{Accion Complementaria Luso-Espanhola} AIC-D-2011-0809, by European FEDER, Spanish MINECO under grant FPA2011-23596 and GVPROMETEO 2010-056. FJB and MN are very grateful for the hospitality of CFTP/IST Lisbon during their visits. MN thanks MINECO for a \emph{Juan de la Cierva} contract.
% % % % % % % % % % % % % % % % % % % % % % % % % % %

\appendix

%%%%%%%%%%%%%%%%%%%%
% % % % % % % % % % % % % % % % % % % % % %
\section{Experimental information\label{sec:expinfo}}
% % % % % % % % % % % % % % % % % % % % % %
%%%%%%%%%%%%%%%%%%%%
This appendix summarizes the experimental information used in the analyses (see \cite{Amhis:2012bh,PDG:2012,Buchalla:2008jp,Antonelli:2009ws}). For simplicity we present them in groups that share some important characteristic: observables related to (dominant) tree level decays, observables involving the mixing in different meson systems like $B_d$'s, $B_s$'s and kaons, rare decays of mesons and precision electroweak information. % Besides listing them, the measured values or the existing bounds that have been used are given.

Tree level observables are those whose extraction from experiment is presumably unaffected by NP effects. These observables include moduli of the CKM elements in the first and second rows. Moduli of third row elements $|\V{tq}|$, $q=d,s,b$, are harder to extract. In fact the only relevant measurement is the one of the ratio of branching fractions $R=\text{Br}(t\to Wb)/\text{Br}(t\to Wq)$, $R=|\V{tb}|^2/(|\V{td}|^2+|\V{ts}|^2+|\V{tb}|^2)$. Notice that, when the mixing matrix is $3\times 3$ unitary, this measurement is rather irrelevant since this unitarity requirement, together with the actual values of $|\V{ub}|$ and $|\V{cb}|$, forces $|\V{tb}|\simeq 1-\mathcal O(10^{-4})$. The picture changes when $3\times 3$ unitarity is not assumed, since significant deviations (much larger than $\mathcal O(10^{-4})$) of $|\V{tb}|$ from $1$ are possible. Finally, the physical phase $\gamma$ (in equation (\ref{eq:angles})), is also a tree level observable. The actual values are collected in table \ref{AP:tab:tree}. The decay $B^+\to\tau^+\nu_\tau$ is also a tree level process\footnote{Notice however that, as it is helicity suppressed and proportional to $|\V{ub}|^2$, NP sizeable contributions may appear in different beyond SM scenarios, but not in our case.}.
\begin{table}[h]
\begin{center}
\begin{tabular}{|c|c|c|c|}
\hline
$|\V{ud}|$ & $0.97425\pm 0.00022$ & $|\V{us}|$ & $0.2252\pm 0.0009$ \\ \hline	
$|\V{cd}|$ & $0.230\pm 0.011$     & $|\V{cs}|$ & $1.023\pm 0.036$   \\ \hline	
$|\V{ub}|$ & $0.00389\pm 0.00044$ & $|\V{cb}|$ & $0.0406\pm 0.0013$ \\ \hline	
$\gamma$   & $(77\pm 14)^\circ$  & $R$ & $0.88\pm 0.07$\\ \hline
Br($B^+\to\tau^+\nu_\tau$) & $(1.68\pm 0.31)\times 10^{-4}$\\ \cline{1-2}
\end{tabular}
\caption{Tree level observables \cite{PDG:2012,Aubert:2007ii,Poluektov:2010wz,Ikado:2006un,Hara:2010dk,Aubert:2007xj,Lees:2012ju}.}\label{AP:tab:tree}
\end{center}
\end{table}

The next set of observables involves the mixings of $B_d$ or $B_s$ mesons (except for $\aslb$, which involves both). We consider time-dependent CP asymmetries $\AJPKs$ and $\AJPP$ (the ``golden'' channel in each system), mass and width differences $\Delta M_{B_d}$, $\Delta\Gamma_d$, and $\Delta M_{B_s}$, $\Delta\Gamma_s$, additional CP asymmetries involving \emph{different} combinations of invariant phases, $\sin(2\bar\alpha)$, $\sin(2\bar\beta+\gamma)$ and $\cos(2\bar\beta)$ (which removes a discrete ambigüity in fixing $2\bar\beta=\sin^{-1}(\AJPKs)$), and, finally, semileptonic asymmetries $\asld$, $\asls$ and $\aslb$. The actual values are collected in table \ref{AP:tab:Bmix}. We will also pay attention to the combination $\asls-\asld$, for which LHCb results are expected.
\begin{table}[h]
\begin{center}
\begin{tabular}{|c|c|c|c|}
\hline
$\AJPKs$ & $0.68\pm 0.02$ & $\Delta M_{B_d}$  & $(0.508\pm 0.004)$\,ps$^{-1}$\\ \hline
$\AJPP$ & $0.002\pm 0.0873$ & $\Delta M_{B_s}$  & $(17.725\pm 0.049)$\,ps$^{-1}$\\ \hline
$\sin(2\bar\alpha)$ & $0.00\pm 0.15$ & $\sin(2\bar\beta+\gamma)$  & $1.0\pm 0.16$\\ \hline
$\cos(2\bar\beta)$& $1.35\pm 0.34$  \\ \hline
$\asld$ & $-0.003\pm 0.0078$ & $\Delta\Gamma_d/\Gamma_d$ & $-0.017\pm 0.021$\\ \hline
$\asls$ & $-0.0017\pm 0.0091$ & $\Delta\Gamma_s$ & $(0.116\pm 0.019)$\,ps$^{-1}$\\ \hline
$\aslb$ & $-0.00787\pm 0.00196$ \\ \cline{1-2}
\end{tabular}
\caption{$B_d$ and $B_s$ mixing-related observables \cite{Amhis:2012bh,PDG:2012,LHCb:2011aa,LHCb:2011ab,Aubert:2009aw,Adachi:2012et}.}\label{AP:tab:Bmix}
\end{center}
\end{table}

Besides the observables involving the mixing, representative rare or suppressed decays are considered. The values in table \ref{AP:tab:Bdecay} include the recent LHCb results used for Br($B_s\to \mu^+\mu^-$) and Br($B_d\to \mu^+\mu^-$) that are playing a key role as the experimental bounds start to probe the SM expected range in the $B_s$ case. The potential for discovery is still quite large in Br($B_d\to \mu^+\mu^-$). The actual values are collected in table \ref{AP:tab:Bdecay}.

\begin{table}[h]
\begin{center}
\begin{tabular}{|c|c|}
\hline
Br($B\to X_s\gamma$) & $(3.56\pm 0.25)\times 10^{-4}$\\ \hline
Br($B\to X_s\mu^+\mu^-$) & $(1.60\pm 0.51)\times 10^{-6}$\\ \hline

Br($B_s\to \mu^+\mu^-$) & $(0\begin{smallmatrix}+2.25\\ --\end{smallmatrix})\times 10^{-9}$\\ \hline
Br($B_d\to \mu^+\mu^-$) & $(0\begin{smallmatrix}+5.15\\ --\end{smallmatrix})\times 10^{-10}$\\ \hline
\end{tabular}
\caption{$B_d$ and $B_s$ rare decays \cite{Amhis:2012bh,PDG:2012,Aad:2012pn,Chatrchyan:2011kr,Chatrchyan:2012rg,Aaij:2012ac,LHCb:2011ac}.}\label{AP:tab:Bdecay}
\end{center}
\end{table}
Moving from $B$ mesons to kaons, representative observables that have to be considered address CP violation in \KKmix\, mixing through $\epsilon_K$ and $\varepsilon^\prime/\epsilon_K$, and rare decays $K^+\to\pi^+\nu\bar\nu$, $K_L\to\pi^0\nu\bar\nu$ and $K_L\to\mu^+\mu^-$. It is worth pointing that the experimental bound on $K_L\to\pi^0\nu\bar\nu$ is overseeded by a model independent bound, Br($K_L\to\pi^0\nu\bar\nu$) $<a\,$ Br($K^+\to\pi^+\nu\bar\nu$) with $a\simeq 4.4$ (see \cite{Grossman:1997sk}). $K_L\to\mu^+\mu^-$ is also expected to play a significant role since the absorptive part of the rate receives a contribution from an intermediate $\gamma\gamma$ state that almost saturates the rate. The actual values are collected in table \ref{AP:tab:kaons}.
\begin{table}[h]
\begin{center}
\begin{tabular}{|c|c|c|c|}
\hline
$\epsilon_K$ & $(2.228\pm 0.011)\times 10^{-3}$\\ \hline
$\varepsilon^\prime/\epsilon_K$ & $(1.67\pm 0.16)\times 10^{-3}$\\ \hline
Br($K^+\to\pi^+\nu\bar\nu$) & $(1.73\pm 1.05)\times 10^{-10}$\\ \hline
Br($K_L\to\pi^0\nu\bar\nu$) & $<\mathcal O(10^{-8})$\\ \hline
Br($K_L\to\mu^+\mu^-$) & $(6.84\pm 0.11)\times 10^{-9}$\\ \hline
\end{tabular}
\caption{Kaon mixing and rare decays \cite{AlaviHarati:1999xp,Ambrose:2000gj,Ahn:2007cd,Artamonov:2008qb,Abouzaid:2010ny,Cirigliano:2011ny}.}\label{AP:tab:kaons}
\end{center}
\end{table}
%\end{table}

For charm-related observables, we have  \DDmix\ mixing, for which we consider \cite{Amhis:2012bh,PDG:2012}
\[
x_D=(0.8\begin{smallmatrix}+0.2\\ --\end{smallmatrix})\times 10^{-2}\ .
\]
It is used to bound the short-distance contribution mediated by flavour changing $Z$-couplings. Notice that the experimental measurement is $x_D=(0.8\pm 0.1)\times 10^{-2}$. As in general \DDmix\ mixing may receive substancial long-distance contributions, but the picture is quantitatively unclear, we content ourselves by requiring that the short-distance contribution does not overproduce this mixing (because in such a case one is implicitly resorting to long-distance contributions \emph{and} a substantial cancellation to keep the theoretical expectation in place). Attention is also paid to the rare decay $D^0\to\mu^+\mu^-$, as GIM suppression yields a branching ratio below $10^{-12}$ in the SM, while current bounds are at the $10^{-7}$ level an rates $\mathcal O(10^{-10})$ are obtainable here.

To complete the ``menu'' of relevant constraints, electroweak precision data has to be considered. We do so through the oblique parameters $S$, $T$, and $U$ \cite{Peskin:1991sw}. The $T$ parameter is the most relevant one, while $S$ is marginal and $U$ negligible in the analysis. Their experimental values (with respect to the SM reference values), are \cite{Flacher:2008zq}
\[
\Delta S=0.02\pm 0.11\,,\qquad\Delta T=0.05\pm 0.12\ ,
\]
with a correlation coefficient $0.879$.\newline
\noindent The experimental values of the different observables quoted along this section are considered to follow gaussian profiles when the uncertainty is quoted in symmetric form, $\mu_X\pm\sigma_X$. This is extended to bounds by modelling them with gaussian profiles too, with central values equal to zero. The only observable not following that prescription is $x_D$, for which no ``sigma'' is quoted for values below $0.8\times 10^{-2}$. As explained above, this only reflects the fact that we model values $x_D\leq 0.8\times 10^{-2}$ to be equally acceptable while a probability toll is paid for values $x_D> 0.8\times 10^{-2}$.

%  Br($K^0\to\pi^0\nu\bar\nu$) & --
Let us comment that, on the theoretical side, several predictions have a limited accuracy that cannot be neglected; this is the case, for example, of some QCD corrections and of several results produced through lattice computations \cite{Colangelo:2010et,Dimopoulos:2011gx,Bazavov:2011aa,McNeile:2011ng,Na:2012kp}. They have associated uncertainties which are fundamental in the analyses. In particular some of the most important ones are quoted in table \ref{AP:tab:lattice}.
\begin{table}[h]
\begin{center}
\begin{tabular}{|c|c|c|c|}
\hline
$\xi=\frac{f_{B_s}\sqrt{B_{B_s}}}{f_{B_d}\sqrt{B_{B_d}}}$ & $1.243\pm 0.028$ &
$f_{B_s}\sqrt{B_{B_s}}$ & $(275\pm 13)$ MeV\\ \hline
$f_{B_s}$ & $(238.8\pm 9.5)$ MeV & $f_{B_s}/f_{B_d}$ & $1.209\pm 0.024$\\ \hline
$B_K$ & $0.72\pm 0.04$ & $f_D$ & $(205\pm 9)$ MeV\\ \hline
\end{tabular}
\caption{Most relevant hadronic input.}\label{AP:tab:lattice}
\end{center}
\end{table}
%%%%%%%%%%%%%%%%%%%%
\section{Numerics\label{sec:numerics}}
%%%%%%%%%%%%%%%%%%%%
In this appendix we explain the numerical procedure underlying the analyses and briefly comment on statistics and the interpretation of the plots presented in the different sections.

Considering the large amount of observables that are of potential interest in, (1), constraining the model or, (2), exploring the possibility of having predictions experimentally distinguishable from SM prospects, an analytic approach to use this information may be interesting for some particular scope, but falls short of the mark when the constraining power of different observables (1) is not clear, (2) changes in different regions of parameter space or (3) when the number of available parameters is large. 

The strategy to deal with these difficulties relies on two cornerstones: considering enough observables to overconstrain the new mixing matrix elements, and using numerical methods that allow an efficient exploration of the much larger parameter space that is available and drawing conclusions from it. The first ingredient is at the heart of the selection of the set of observables listed in appendix \ref{sec:expinfo}. Let us elaborate on the second one. 
First of all: which are the (new) parameters to be used? They are
\begin{itemize}
\item the new mass eigenvalue $\mT$,
\item the parameters necessary to describe the extended CKM mixing matrix. As it is a submatrix of a $4\times 4$ unitary matrix, we can resort to appropriate generalizations \cite{Botella:1985gb} of the standard Chau \& Keung \cite{Chau:1984fp} parameterization of the $3\times 3$ unitary standard CKM matrix. This requires introducing \emph{three} new ``inter-family'' mixing angles $\theta_{14}$, $\theta_{24}$, $\theta_{34}$ (in addition to the usual $\theta_{12}$, $\theta_{13}$ and $\theta_{23}$) and \emph{two} new complex phases $\delta_{14}$ and $\delta_{24}$ (in addition to the usual $\delta_{13}$).
\end{itemize}
We are thus left with \emph{six} new, beyond SM, parameters; a point in parameter space is essentially given by the values of $\{$ $\theta_{12}$, $\theta_{13}$, $\theta_{23}$, $\theta_{14}$, $\theta_{24}$, $\theta_{34}$, $\delta_{13}$, $\theta_{14}$, $\theta_{24}$, $\mT$ $\}$. Mixing angles and phases cover the usual ranges; for the new mass eigenstate, values above $2.5$ TeV are not explored since for those values decoupling through suppressed mixings is enforced and only minute effects are produced.
%}
For any given point in parameter space, we can then compute the values of all the considered observables. Comparison of this set of predictions \cite{Lavoura:1992np,Barenboim:1997pf,Barenboim:1997qx,Chang:2000gz,Aoki:2000ze,AguilarSaavedra:2002kr} with the corresponding measurements allows us to define, as usual, a \emph{likelihood} function $\mathcal L$ (or equivalently a $\chi^2$ function $\chi^2=-2\ln\mathcal L$) everywhere in parameter space. This function has all the information we could be interested in: how good is the model to reproduce the experimental values of all the considered observables at any given point in parameter space, hence over all the parameter space. The difficulty is that this function lives in a parameter space with too many dimensions to be easily grasped. This poses two problems: how do we explore this high-dimensional parameter space and how do we reduce the information to something that we can handle. The first one is a matter of the numerical methods available while the second one is a question of statistics.

\underline{Numerical procedure}. 
When dealing with such a number of observables and parameters, simple MonteCarlos are hardly appropriate for the task. The method of choice are Markov-chain MonteCarlos, in particular improved versions of the ``classical'' Metropolis algorithm. They allow an efficient exploration of parameter space: as a tool for sampling the multidimensional likelihood function, obtaining probability distribution functions for a number of interesting quantities is almost straightforward; in addition, through appropriate modifications, likelihood profiles are equally obtainable.

\underline{Statistics}. Reducing the information of $\mathcal L$ in the $n$-dimensional parameter space to zero, one or two-dimensional functions is a non-trivial task. There are two main ``schools'', bayesians and frequentists. Without entering the polemic arena of the respective advantages and drawbacks we present frequentist results: $\Delta\chi^2$ profiles for individual quantities (from which confidence intervals can be easily obtained) and canonical 68\%, 95\% and 99\% CL regions in two-dimensional plots.

Beside the statistical treatment adopted, one may wonder if the overall goodness of a fit with the \VL model compares well with the SM one: for that to be, the increase in the number of parameters of the model should improve significantly on the SM result. The quantity of interest for that comparison is the value of $\chi^2$ per degree of freedom for each fit. While for the SM we have $1.53$ for 27 degrees of freedom, for the \VL model we have $1.56$ for 21 degrees of freedom; the goodness of both fits is similar, i.e. the increase in the number of parameters in the \VL really yields better fits to the existing data.

One last comment closes this appendix. As far as the formulas we have used for the \VL model, we have to stress that we have included for the first time in this type of analysis the right decoupling behaviour of the singlet up vector-like quark $T$ \cite{Botella:2012zbs,Nardi:1995fq,Vysotsky:2006fx,Kopnin:2008ca,Picek:2008dd}. In fact it amounts to use always the exact formula for $V-A$ coupling of the $Z$ to up fermions $(\CKM\,\CKM^\dagger)_{ij}=\delta_{ij}-U_{i4}^{\phantom{\ast}}U_{j4}^{\ast}$ and never use the approximation $(\CKM\,\CKM^\dagger)_{ij}=\delta_{ij}$, even if from a numerical point of view it may look irrelevant in some cases.

%\providecommand{\href}[2]{#2}\begingroup\raggedright\begin{thebibliography}{10%0}

%\endgroup

\end{document}